\documentclass[journal]{IEEEtran}
\usepackage[numbers,sort&compress,square]{natbib}
\usepackage{comment}
\usepackage{textcomp}
\usepackage{gensymb}
\usepackage[pdftex]{graphicx}
\usepackage{amsmath}
\usepackage{amssymb}
\usepackage{array}
\usepackage{multirow}
\usepackage[acronym,toc,shortcuts]{glossaries}
\usepackage{commath}
\usepackage{arydshln}
\usepackage{mathtools}
\usepackage{stfloats}
\usepackage{xcolor}
\usepackage{url}
\usepackage{subcaption}
\usepackage[T1]{fontenc}
\usepackage{enumitem}
\usepackage{ragged2e}

\newcommand{\quotes}[1]{``#1''}

\DeclarePairedDelimiter{\floor}{\lfloor}{\rfloor}

\newacronym{GNSS}{GNSS}{Global Navigation Satellite System}
\newacronym{DGPS}{DGPS}{Differential Global Positioning System}
\newacronym{VLP}{VLP}{visible light positioning}
\newacronym{VLC}{VLC}{visible light communication}
\newacronym{RF}{RF}{radio frequency}
\newacronym{DSRC}{DSRC}{dedicated short range communications}
\newacronym{TX}{TX}{transmitters}
\newacronym{RX}{RX}{receiver}
\newacronym{ToA}{ToA}{time-of-arrival}
\newacronym{TDoA}{TDoA}{time-difference-of-arrival}
\newacronym{PDoA}{PDoA}{phase-difference-of-arrival}
\newacronym{RToF}{RToF}{roundtrip-time-of-flight}
\newacronym{RSS}{RSS}{received signal strength}
\newacronym{AoA}{AoA}{angle-of-arrival}
\newacronym{COTS}{COTS}{commercial off-the-shelf}
\newacronym{LoS}{LoS}{line-of-sight}
\newacronym{MUSIC}{MUSIC}{multiple signal classification}
\newacronym{FoV}{FoV}{field-of-view}
\newacronym{PD}{PD}{photodiode}
\newacronym{QRX}{QRX}{quadrant-photodiode-based AoA-sensing VLC RX}
\newacronym{VLP-VPE}{VLP-VPE}{VLP-based vehicle pose estimation}
\newacronym{MIMO}{MIMO}{multiple-input-multiple-output}
\newacronym{QPD}{QPD}{quadrant-photodiode}
\newacronym{BFSK}{BFSK}{binary frequency shift keying}
\newacronym{AWGN}{AWGN}{additive white gaussian noise}

\begin{document}
% at, but, by, for, in, nor, of, on, or, the, to and up, 
\title{Visible Light Communication based Vehicle Localization for Collision Avoidance and Platooning}


\author{Burak~Soner,~\IEEEmembership{Student Member,~IEEE,}
        and~Sinem~Coleri,~\IEEEmembership{Senior Member,~IEEE,}% <-this % stops a space
\thanks{Burak Soner and Sinem Coleri are with the Department of Electrical and Electronics Engineering, Koc University, 34450 Istanbul, Turkey (e-mail: bsoner16@ku.edu.tr, scoleri@ku.edu.tr). Burak Soner is also with Koc University Ford Otosan Automotive Technologies Laboratory (KUFOTAL), Istanbul, Turkey. The authors acknowledge the support of Ford Otosan and the Scientific and Technological Research Council of Turkey EU CHIST-ERA grant \# 119E350.}}% <-this % stops a space

\markboth{placeholder}%
{Shell \MakeLowercase{\textit{et al.}}: Bare Demo of IEEEtran.cls for IEEE Journals}



\maketitle

\begin{abstract}
Collision avoidance and platooning applications require vehicle localization at cm-level accuracy and at least 50 Hz rate for full autonomy. The RADAR/LIDAR and camera based methods currently used for vehicle localization do not satisfy these requirements, necessitating complementary technologies. Visible light positioning (VLP) is a highly suitable complementary technology due to its high accuracy and high rate, exploiting the line-of-sight propagation feature of the visible light communication (VLC) signals from LED head/tail lights. However, existing vehicular VLP algorithms impose restrictive requirements, e.g., use of high-bandwidth circuits, road-side lights and certain VLC modulation strategies, and work for limited relative vehicle orientations, thus, are not feasible for general use. This paper proposes a VLC-based vehicle localization method that eliminates these restrictive requirements by a novel VLC receiver design and associated vehicular VLP algorithm. The VLC receiver, named QRX, is low-cost/size, and enables high-rate VLC and high-accuracy angle-of-arrival (AoA) measurement, simultaneously, via the usage of a quadrant photodiode. The VLP algorithm estimates the positions of two head/tail light VLC transmitters (TX) on a neighbouring vehicle by using AoA measurements from two QRXs for localization. The algorithm is theoretically analyzed by deriving its Cramer-Rao lower bound on positioning accuracy, and simulated localization performance is evaluated under realistic platooning and collision avoidance scenarios. Results demonstrate that the proposed method performs at cm-level accuracy and up to 250 Hz rate within a 10 m range under realistic harsh road and channel conditions, demonstrating its eligibility for collision avoidance and safe platooning.
\end{abstract}

\begin{IEEEkeywords}
autonomous vehicles, platooning, collision avoidance, vehicle localization, visible light communication.
\end{IEEEkeywords}

% \fi
\IEEEpeerreviewmaketitle

\section{Introduction}
\vspace{3mm}
\IEEEPARstart{A}{utomotive} research is currently heavily oriented towards vehicular automation and autonomy, and the foremost objective is improving driving safety and efficiency \cite{ford_trust}. The annual traffic accident report published by the Federal Statistical Office of Germany (DESTATIS) \cite{destatis} shows that 63\% of traffic accidents are vehicle-to-vehicle collisions, demonstrating the importance of collision avoidance systems and safe platooning for future automated/autonomous vehicle safety concepts \cite{benefits_of_fcw}. Collision avoidance and platooning systems require relative vehicle localization with at least 50 Hz rate and cm-level accuracy with high reliability and availability under harsh road conditions \cite{caveney_coopVehSfty,shladover_vehPosAccReqs,depontemuller_survey}.

Current sensor-based methods, which are readily being used for less demanding conventional autonomous driving tasks, fail to meet the rate and accuracy requirements of collision avoidance and platooning systems \cite{depontemuller_survey}. Differential Global Positioning System (DGPS), which is used for global self-localization, allows vehicles to also cooperatively localize each other. However, this sensor provides only meter-level accuracy at less than 20 Hz rate \cite{gpsRtkAnalyz}, and cannot provide accurate localization since it regards vehicles as point objects. Alternatively, RADAR/LIDAR \cite{lidar_ref} and camera-based methods \cite{stixel}, which are used for the localization of non-vehicle objects on the road, can be used for vehicle localization at up to cm-level accuracy. However, these methods are limited to less than 50 Hz rate since they require scanning, locating and labelling millions of points/pixels for object localization \cite{3dvehmdl,visionOnlyLcl}. On the other hand, communication-based positioning methods, which promise estimation of antenna positions at cm-level accuracy and greater than 50 Hz rate, can be extended for vehicle localization, enabling fully autonomous collision avoidance and safe platooning, and complementing the existing autonomous driving system for higher safety and driving efficiency.

Communication-based positioning methods in the vehicular domain mainly employ radio frequency (RF) technologies such as cellular and dedicated short range communications (DSRC) \cite{cellularDsrcInterwork}, and VLC technologies \cite{balico_survey, smartAutoLighting, gurbilek2019location, revistedi_vlcrf}. Cellular-based positioning methods either utilize location-fingerprinted received signal strength (RSS) measurements or apply triangulation via physical system parameters such as time-of-arrival (ToA), time-difference-of-arrival (TDoA) and angle-of-arrival (AoA) \cite{liu_wiComPosSurvey}. However, these methods rely on tight synchronization between base stations and the mobile terminal, with limited accuracy due to excessive multi-path interference. Although pilot-based synchronization methods \cite{gustafsson} and estimators like multiple signal classification (MUSIC) \cite{vehaoa} somewhat mitigate these problems, overall cellular-based positioning accuracy is worse than 10 m in practical scenarios \cite{practical_cellular, hybrid_globecom}. DSRC also suffers from similar issues; while roundtrip-time-of-flight (RToF) methods successfully mitigate synchronization issues \cite{alam_52, alam_53}, the best reported accuracy is around 1-10 m \cite{NAlam_cooperative}, still worse than the cm-level requirement. As an alternative, vehicular VLP methods based on line-of-sight (LoS) VLC signals from automotive LED head/tail lights fundamentally promise cm-level accuracy at near-kHz rate \cite{zhuangSurvey_posLed, smartAutoLighting}. However, this promise is not fulfilled in practice since existing methods impose restrictive requirements such as the use of high-bandwidth circuits or road-side lights, and constraints on the VLC subsystem such as wasting communication sub-carriers for positioning \cite{revistedi_vlcp}, and they only work for limited relative vehicle orientations.

Previous works in vehicular VLP use TDoA \cite{bai_tdoa}, phase-difference-of-arrival (PDoA) \cite{roberts_pdoa, smartAutoLighting}, RToF \cite{bechadergueConf_rtofPdoa, bechadergueThs_rtofPdoa} and AoA \cite{mfkeskin, yamazato2014_camAoA, yamazato2014_hispeed} approaches. In \cite{bai_tdoa}, vehicles utilize TDoA of the VLC signals from globally localized traffic lights to two on-board photodiodes for estimating their own global positions. This self-localization method can trivially be extended for relative localization through vehicles exchanging their global positions, but accuracy is around 1 m for realistic conditions and the method has low availability since it restrictively requires the presence of localized traffic lights. In \cite{smartAutoLighting} and \cite{roberts_pdoa}, the PDoA of head/tail light VLC signals to two photodiodes on a vehicle provides cm-level positioning. However, the method restrictively assumes that the vehicles are oriented parallel to each other and requires very high frequency constant tones (10-50 MHz), which cannot practically be transmitted to useful distances with automotive LEDs \cite{caileanSurvey_vlcAutoChlgs}. Recently, \cite{bechadergueConf_rtofPdoa} and \cite{bechadergueThs_rtofPdoa} have achieved positioning with cm-level longitudinal accuracy at kHz rate based on the RToF of a VLC message between two vehicles. However, the method suffers from low lateral accuracy due to the sensitivity of the underlying geometry and restrictively requires special high-bandwidth circuits, hence, cannot provide cm-level vehicle localization under feasible operation conditions. A VLP method that is devoid of such restrictive requirements is necessary for vehicle localization in collision avoidance and platooning applications.

AoA-based VLP methods promise high accuracy without imposing restrictive requirements such as limited vehicle orientations, presence of road-side/traffic lights, and high-bandwidth circuit and VLC modulation constraints \cite{mfkeskin}. However, their vehicular implementations have so far been limited to camera-VLC based methods. Camera-VLC approaches are not suitable since they either provide very low (\textless kbps \cite{caileanSurvey_vlcAutoChlgs}) communication rates \cite{yamazato2014_camAoA} or require costly high-frame-rate cameras \cite{yamazato2014_hispeed}, which beats the purpose of VLP complementing sensor-based methods. Therefore, a low-cost and small-size photodiode-based VLC receiver design that can provide high-rate VLC and high-accuracy, high-resolution and high-rate AoA measurement, is needed.

Existing photodiode-based VLC receiver designs that can be used for AoA measurement fall into four main categories \cite{c_he_ADA_rx}: aperture-based, lens-based, prism-based and tilted-photodiode-based designs. Tilted-photodiode designs \cite{optimal_aoa, cornercuberx, sphericalRX, ir_bizimgeo} and special prism-based designs \cite{wang2014prism} are not suitable for vehicular use since they are not low-cost/size and typically provide limited AoA resolution (i.e., coarsely quantized AoA intervals as estimates). Aperture-based designs using commercially available low-cost/size quadrant-photodiodes (originally proposed for angular diversity in multiple-input-multiple-output (MIMO) indoor VLC \cite{esteve_qadaish, qada_origins, kahn_wiIR}) can be used for accurate and resolute AoA measurement but the aperture limits the field-of-view (FoV). Using an imaging architecture, e.g., a hemispherical lens rather than an aperture, provides larger FoV \cite{wang_hemisphericalAnalyz}. Such quadrant-photodiode-based imaging designs, traditionally used for laser target tracking \cite{carbonneau1986optical} and transceiver pointing \cite{soner_IAT}, are also promising for AoA measurement. A low-cost/size realization of this architecture with commercial-off-the-shelf (COTS) components would enable the restriction-free AoA-based high accuracy and rate vehicular VLP method.

In this paper, we propose a VLC/VLP-based vehicle localization method that uses only two on-board AoA-sensing receivers for obtaining the relative 2D location of a transmitting vehicle. First, we provide a novel low-cost/size VLC receiver (RX) design which enables high-rate VLC and high-accuracy, high-resolution and high-rate AoA measurement, simultaneously; we call this design \quotes{QRX}. This design enables the first practical vehicular implementation of an AoA-based VLP method for localization at cm-level accuracy and higher than 50 Hz rate. Two QRX units located at the head/tail lights measure the AoA from two VLC head/tail light transmitters (TX) on the target vehicle for estimating the positions of the TXs separately via triangulation (i.e., dual AoA measurements and the inter-QRX distance defines a triangle). The two TX positions, which are known to be on the edges of the front/rear faces of the target vehicle, sufficiently define the vehicle location. This work extends our previous related work, where a single QRX is considered for VLP but the QRX design is not presented and the VLP method necessitates the target vehicle to disseminate its heading and speed information via VLC \cite{soner_pimrc}. This paper provides design details for the QRX, and the method proposed in this paper does not require any such co-operation from the transmitting vehicle since it applies triangulation directly with two on-board QRXs. The main contributions of this paper are given as follows: 

\vspace{0mm}
\begin{itemize}
	\item We present a novel low-cost/size VLC receiver design (QRX), which uses only COTS components and enables high-rate VLC and high-accuracy, high-resolution and high-rate AoA measurement, simultaneously, for the first time in the literature. The design is a quadrant-photodiode-based imaging receiver similar to \cite{wang_hemisphericalAnalyz} but specifically designed for AoA-based vehicular VLP.
	\vspace{2mm}
	\item We propose an AoA-based vehicular VLP algorithm that uses two of the designed QRXs and promises localization at cm-level accuracy and greater than 50 Hz rate without imposing any restrictive requirements like the use of road-side lights, high-bandwidth circuits and certain VLC modulation strategies, for the first time in the literature.
	\vspace{2mm}
	\item We derive the Cramer-Rao lower bound (CRLB) on positioning accuracy for the dual-AoA vehicular VLP geometry used by our algorithm, for the first time in the literature. Since the bound is associated with the vehicular VLP geometry used by the algorithm, it applies to all vehicular localization methods that use the same dual-AoA geometry but different AoA measurement procedures (e.g., the procedure in \cite{steendam_sqrarray} can be utilized).
	\vspace{2mm}
	\item We evaluate the derived CRLB and run extensive simulations for the proposed method under realistic driving scenarios that consider different weather (i.e., clear, rainy and foggy) and ambient light conditions (i.e., night-time and day-time). The results demonstrate cm-level accuracy and greater than 50 Hz rate under majority of these comprehensive scenarios, conclusively proving the eligibility of VLC-based vehicle localization for use in collision avoidance and safe platooning applications for the first time in the literature.
\end{itemize}
\vspace{0mm}

The rest of the paper is organized as follows. Section II presents the mathematical model of the vehicular VLC/VLP system and defines the problem of vehicle localization using AoA measurements from received VLC signals. Section III presents the first part of our proposed VLC-based vehicle localization method, which is the novel low-cost/size QRX design and the associated AoA measurement procedure. Section IV presents the second part, the AoA-based VLP algorithm for vehicle localization, and derives the CRLB on positioning accuracy for the underlying geometry. Section V demonstrates the performance of the proposed method at the required accuracy and rate for collision avoidance and platooning by both evaluating the derived theoretical CRLB on positioning accuracy and extensive simulations. A custom MATLAB©-based simulator is used for evaluations under realistic road and VLC channel conditions. Section VI concludes the paper.


\section{System Model and Problem Definition}
%\vspace{2mm}

This section first presents the mathematical model governing the vehicular VLC/VLP system, and then defines the VLC-based vehicle localization problem in collision avoidance and platooning considering AoA-based VLP.

\vspace{-2mm}
\subsection{System Model}
%\vspace{2mm}

The model considers the following assumptions (\textit{A\#}): 

\vspace{1mm}
\begin{itemize}
	\item \textit{A1}: Vehicles drive on piecewise-flat roads, i.e., neighbouring vehicles share flat road sections and have no pitch angle difference between them. This assumption, which allows to define the vehicle localization problem in 2D, is reasonably valid for collision avoidance and platooning scenarios since these scenarios consider vehicles within 1 to 20 m distance of each other driving at speeds greater than 30 km/h \cite{sartre, benefits_of_fcw, geoRoad2D}.
	\vspace{2mm}
	\item \textit{A2}: Vehicles contain two VLC units each on the edges of both front and rear faces, i.e., on LED head/tail lights. Each VLC unit utilizes its LEDs as the TX, contains one AoA-sensing RX, and sustains reliable LoS VLC \cite{vlc_channel_lee, vlc_channel_wnl} with other units in its FoV.
	\vspace{2mm}
	\item \textit{A3}: VLC TX units are assumed to be point sources from the RX perspective, i.e., received optical power obeys the inverse square law with respect to distance. This assumption is reasonably valid for the considered scenarios since the 1 m minimum distance between the vehicles (\textit{A1}) is larger than the photometric distance for vehicle LED lights, which is at most 50 cm \cite{iala_farfield, autoOptikDist_farfield, 5to15_farfield}.
	\vspace{2mm}
	\item \textit{A4}: Transmissions by the VLC units do not interfere. This can be achieved through the design of a medium access control mechanism for the network containing VLC units, on both the same vehicle and neighbouring vehicles, such that no two units transmit at the same time on the same frequency band when their lines-of-sight are towards the same RX \cite{mac_layer_dcf, mac_layer_survey}. Note that such a mechanism does not necessitate explicit identification of the units.
\end{itemize}
\vspace{1mm}

\begin{figure}[!t]
	\centering
	\includegraphics[width=.93\linewidth]{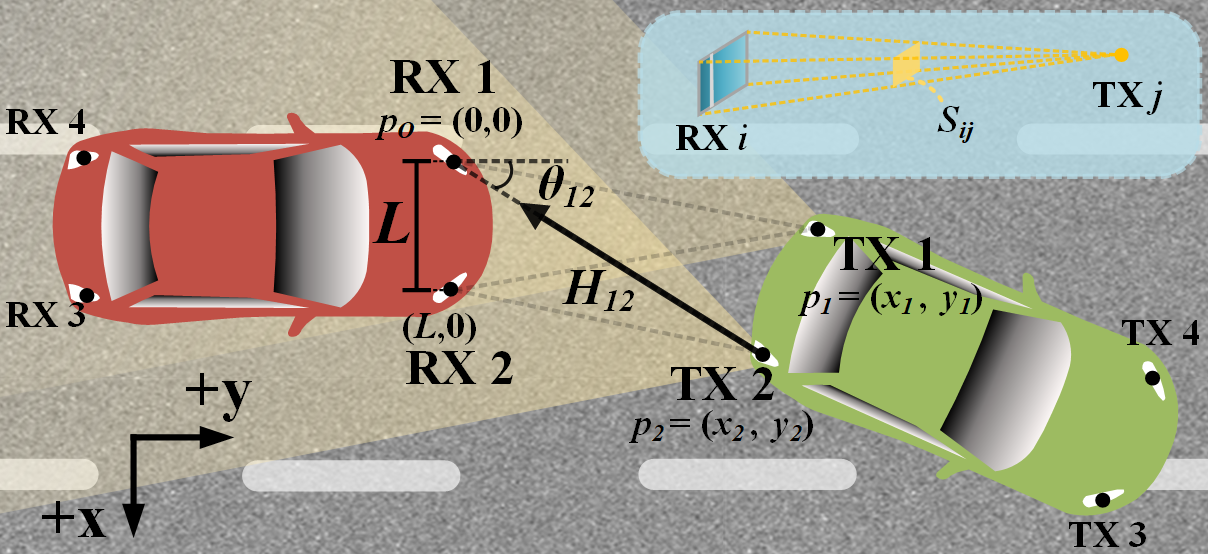}
	\vspace{0mm}
	\caption{System model. $p_{1} \!= \!(x_{1}, y_{1})$ and $p_{2} \! = \! (x_{2}, y_{2})$ are target (green) TX 2D positions relative to the ego vehicle (red) origin $p_{O} \! = \! (0,0)$. $H_{ij}$ is the channel gain and $\theta_{ij}$ is the AoA from TX $j$ to RX $i$, where $i,j \in \{1,2,3,4\}$, and the case for $i$=1 and $j$=2 is shown as example. $S_{ij}$ is the solid angle subtended by RX $i$ with respect to TX $j$ and $L$ is the RX separation.}
	\label{sys_mdl}
	\vspace{-5mm}
\end{figure}

Based on these assumptions, the mathematical model of the received VLC signals is as follows:

\vspace{-2mm}
\begin{equation}
	\label{rx_model_1}
	r_{i} = \sum\limits_{j} r_{ij}  + \mu_{i} ~~,~~ r_{ij} = H_{ij}~ s_{j}~~,
\end{equation}
\vspace{-2mm}

\noindent where $i, j \in \{1,2,3,4\}$ are indices for RXs and TXs on vehicle lights respectively, $r_{i}$ is the total received photocurrent signal, $s_{j}$ is the transmitted photocurrent signal, $r_{ij}$ is the contribution of $s_{j}$ to $r_{i}$, $\mu_{i}$ is the photocurrent additive white Gaussian noise (AWGN), and $H_{ij}$ is the geometric channel gain from TX $j$ to RX $i$. Since no two TX units can transmit on the same frequency band at the same time as per assumption (\textit{A4}), signals $r_{ij}$ can be extracted from $r_{i}$ by band-pass filtering $r_{i}$ for the respective bands occupied by $s_j$. Under the point-source TX assumption (\textit{A3}), $H_{ij}$ is expressed as:

\begin{subequations}
\vspace{-1mm}
\begin{equation}
	\label{rx_model_2}
	H_{ij} = \lambda_{i}(\theta_{ij}) \iint \limits_{S_{ij}} \gamma_{j} \rho_{j}(S)~ dS
\end{equation}
\vspace{-3mm}
\begin{equation}
	\label{rx_model_aoa}
	S_{ij} \propto \frac{A_i \cos(\theta_{ij})}{\sqrt{{x_{ij}}^2 \! + \! {y_{ij}}^2}}~~,~~ \theta_{ij} = \arctan \left( \frac{x_{ij}}{y_{ij}} \right)~~,
\end{equation}
\vspace{-1mm}
\end{subequations}

\noindent where $\rho_{j}(S)$ is the normalized positive-definite beam pattern and $\gamma_{j}$ is electrical-to-optical gain for TX $j$, $S_{ij}$ is the solid angle subtended by the active area of RX $i$ with respect to TX $j$ \cite{solid_angle}, $A_{i}$ is the active area and $\lambda_{i}$ is the AoA-dependent sensitivity of the photodiode in RX $i$, and $\theta_{ij}$ is the AoA from TX $j$ to RX $i$. $(x_{ij},~y_{ij})$ denotes the 2D location of TX $j$ relative to RX $i$ for system description. However, since $(x_{2j},~y_{2j})$ is equal to $(x_{1j}-L,~y_{1j})$ by definition, where $L$ is the RX separation, the $i$ subscript in $(x_{ij},~y_{ij})$ is dropped in the rest of the paper, i.e., $p_j = (x_{j},~y_{j}) \! = \! (x_{1j},~y_{1j})$ for TX $j$, $j \in \{1,2\}$, as shown in Fig. \ref{sys_mdl}. Eqn. (\ref{rx_model_2}) can be converted to a closed-form expression when a Lambertian model is assumed \cite{crlbApertureRX}. Nevertheless, the more general model is provided here since regulation-compliant automotive LED lights are not strictly Lambertian emitters. $\mu_{i}$ is composed of shot noise on the RX photodetector (PD, the COTS p-i-n type is assumed) and thermal noise on the FET-based front-end transimpedance amplifier (TIA) that converts $r_{i}$ to a voltage signal for processing \cite{smith1980receiver}. Therefore, $\mu_{i}$ is zero-mean and has variance $ \sigma_{\mu_{i}}^2 = \sigma_{shot_{i}}^2 + \sigma_{thm_{i}}^2 $~, where

\begin{subequations}
	\vspace{-1mm}
	\begin{equation}
		\label{rx_noise_shot}
		\sigma_{shot_{i}}^2 = 2 q \gamma_{i} P_{r,i} B_{i} + 2 q I_{bg,i} I_{B2} B_{i}  
	\end{equation}
	\vspace{-3mm}
	\begin{equation}
		\label{rx_noise_thermal}
		\sigma_{thm_{i}}^2 = 4k T_{i} \left( \frac{1}{R_{F,i}} I_{B2} B_{i} \! + \!  \frac{\left(2 \pi C_{T,i} \right)^2}{g_{m,i}} \Gamma I_{B3} B_{i}^3 \right)~,
	\end{equation}
	\vspace{0mm}
\end{subequations}

\noindent and $q$ is the Coulomb electron charge, $k$ is the Boltzmann constant, $P_{r,i}$ is the received optical signal power, $I_{bg,i}$ is the background illumination current, $B_{i}$ is the front-end bandwidth, $T_{i}$ is the circuit temperature, $R_{F,i}$ is the front-end resistance (i.e., TIA feedback gain term), $C_{T,i}$ is the input capacitance due to the photodiode and the FET and $g_{m,i}$ is the FET transconductance, on RX $i$, and $\Gamma$, $I_{B2}$ and $I_{B3}$ are unitless factors for FET channel noise and noise bandwidth determined by the signal shape \cite{kahn_wiIR}. For a bandwidth-optimal TIA (i.e. proper loop compensation and impedance matching \cite{smith1980receiver}), Eqn. (\ref{rx_noise_thermal}) is typically reorganized by using $R_{F,i} = G/(2 \pi B_{i} C_{T,i})$, where $G$ is the commonly referred \quotes{open-loop voltage gain}, and the front-end circuit gain is independent of transistor parameters, i.e., $R_{F,i}$ determines the transimpedance gain that converts photocurrent $r_i$ to a voltage signal. Popcorn noise due to silicon defects are ignored since modern components are used, and also flicker (i.e., $1/f$ noise) is ignored since VLC operation is not near DC. Furthermore, the finite light propagation time from TX $j$ to RX $i$ is ignored since AoA-based VLP is considered and VLC units are assumed to be synchronized as per assumption (\textit{A2}), and random fluctuations on $H_{ij}$ due atmospheric turbulence on the channel are ignored since automotive LEDs are non-coherent.

The model is depicted in Fig. \ref{sys_mdl}: The red vehicle, termed the \quotes{ego} vehicle, finds the relative position of the TX units on the green vehicle, termed the \quotes{target} vehicle, for relative localization via AoA-based VLP using RX signals. In this paper, TX units are in the target vehicle and RX units are in the ego vehicle, but a vehicle can take on either role since it contains both TX and RX units in both head and tail lights.

%\vspace{-3mm}
\subsection{Problem Definition}

\begin{figure}[!t]
	%\vspace{-5mm}
	\centering
	\begin{subfigure}[t]{.89\linewidth}
		\includegraphics[width=.89\linewidth]{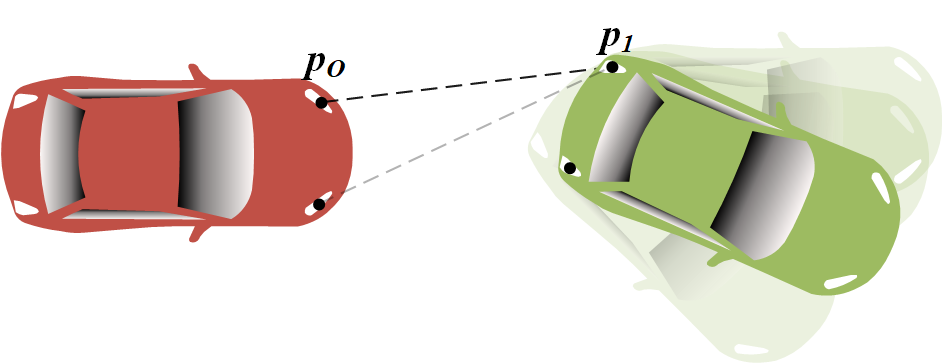}%
		\vspace{-1mm}
		\caption{1 TX position $\rightarrow$ indefinite but bounded location set}
	\end{subfigure}
	\begin{subfigure}[t]{.89\linewidth}
		\vspace{1mm}
		\includegraphics[width=.89\linewidth]{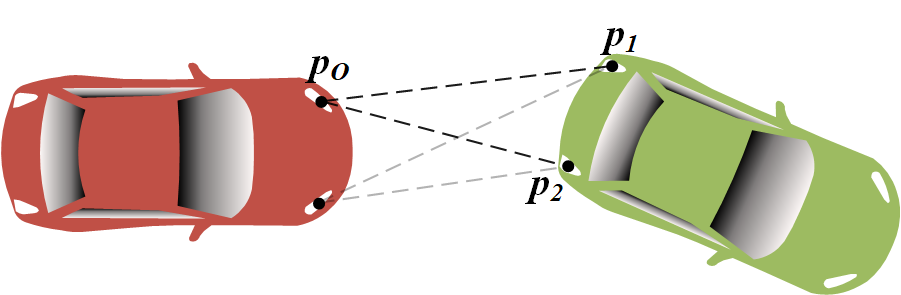}%
		\vspace{-1mm}
		\caption{2 TX positions $\rightarrow$ definitive vehicle localization}
	\end{subfigure}
	\begin{subfigure}[t]{.95\linewidth}
		\vspace{3mm}
		\includegraphics[width=.95\linewidth]{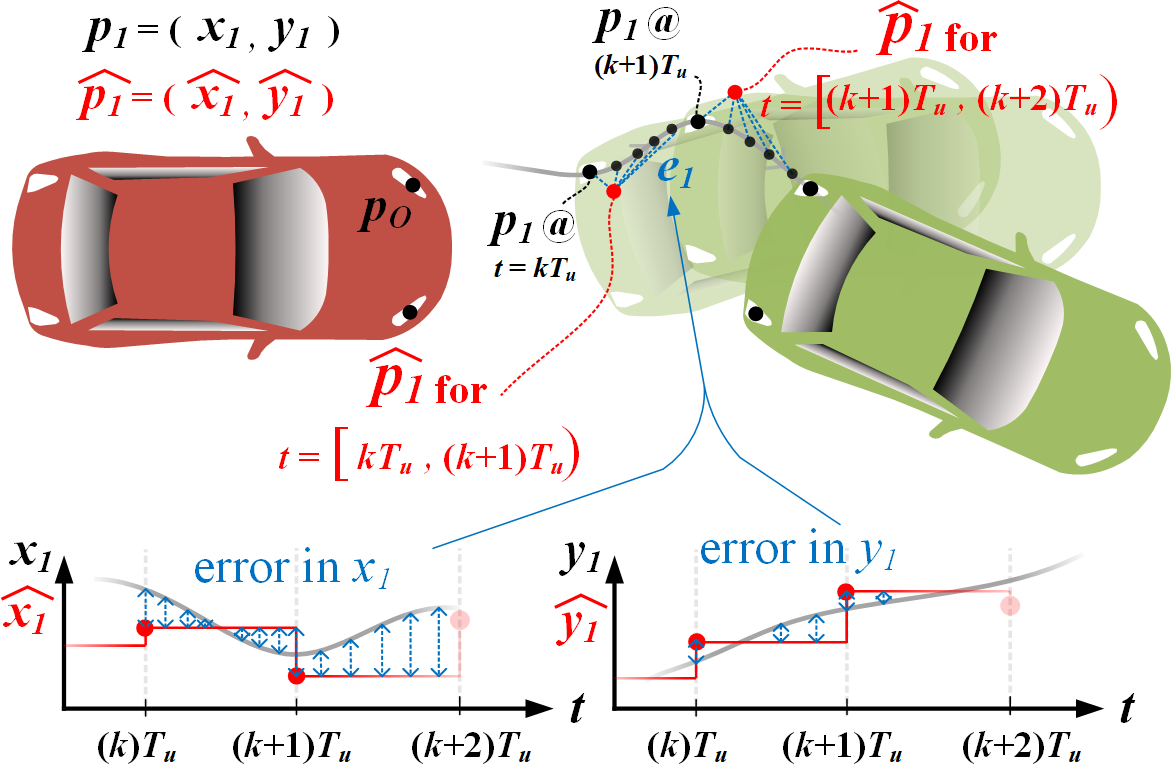}%
		\vspace{2mm}
		\caption{Accuracy and rate jointly determine localization performance}
	\end{subfigure}
	\vspace{1mm}
	\caption{Problem definition. $p_1$, $p_2$ and $\widehat{p_1}$, $\widehat{p_2}$ are actual and estimated positions for TX 1, TX2, respectively, relative to the ego vehicle origin, $p_O$.}
	\vspace{-3mm}
	\label{probdef}
\end{figure}

The ego vehicle needs to find the relative position of two TX units on the target vehicle for determining its relative location as shown in Fig. \ref{probdef}b; although one TX position creates a bounded solution set, it does not define an exact vehicle location, as shown in Fig. \ref{probdef}a. Since the ego vehicle uses noisy RX signals for finding TX positions, the location estimates have finite accuracy and error $e$ is given by the distance between estimated and actual TX positions at a given time: 

\vspace{-2mm}
\begin{equation}
	\label{accuracy}
	e = 
	\left[  
	\begin{matrix}
		e_{1}\\
		e_{2}
	\end{matrix}
	\right] = 
	\left[ 
	\begin{matrix}
		~\sqrt{ \left( x_{1} - \widehat{x_{1}}  \right)^2 +  \left( y_{1} - \widehat{y_{1}}  \right)^2}~\\
		~\sqrt{ \left( x_{2} - \widehat{x_{2}}  \right)^2 +  \left( y_{2} - \widehat{y_{2}}  \right)^2}~
	\end{matrix}
	\right]~,
\end{equation}
\vspace{-1mm}

\noindent where $(\widehat{x_{1}},~\widehat{y_{1}})$ and $(\widehat{x_{2}},~\widehat{y_{2}})$ are estimations for TX 1 and \linebreak TX 2 positions, also called $\widehat{p_1}$ and $\widehat{p_2}$, respectively, and $e_1$ and $e_2$ are the associated errors. The \quotes{cm-level} accuracy requirement therefore denotes that the norm of $e$ should at most be 10 cm at a given time. However, since the vehicles are in continuous relative movement and localization occurs at a finite rate, such a static accuracy definition is not sufficient, and rate should also be considered in determining the localization performance. 

Let $f_u$ be the localization rate and $T_u = 1/f_u$ be the localization update period, i.e., the time between the $k$\textsuperscript{th} and the $(k+1)$\textsuperscript{th} estimate, where $k \in \{0,1,2,...\}$. \linebreak Localization accuracy continuously varies since the relative target location does not stay constant between consecutive estimates; this is illustrated with exaggeration in Fig. \ref{probdef}c. Assuming that an estimate at time $t \! = \! kT_u$ has sufficient accuracy for the location at $t \! = \! kT_u$, $f_u$ needs to be higher than $\nu_u/(10~cm)$ for ensuring cm-level accuracy until the next estimate at $t \! = \! (k \! + \!1)T_u$, where $\nu_u$ is the relative target speed. To satisfy this in most feasible collision avoidance and platooning scenarios, $f_u$ should be at least 50 Hz, constituting the associated requirement for estimation rate \cite{veh_dyn_Hzlevel,	shladover_vehPosAccReqs}.

While AoA-based VLP promises to satisfy these rate and accuracy requirements without imposing the restrictive constraints described earlier \cite{mfkeskin, roadmap_vlp}, its practical realization for vehicle localization faces two main challenges, which constitute the main problem considered in this paper: 

\begin{enumerate}[label={\arabic*)}]
	\vspace{0.5mm}
	\item The AoA-sensing VLC receiver design should be realizable with low-cost, COTS components and needs to provide high-rate VLC and high-accuracy, high-resolution and high-rate AoA measurement, simultaneously, despite adverse road and channel conditions.
	\vspace{2mm}
	\item The VLP algorithm, which uses AoA measurements to find relative TX positions, should be robust against erroneous AoA measurements due to the adverse conditions, and needs to provide vehicle localization with cm-level accuracy and at least 50 Hz rate.
	\vspace{0.5mm}
\end{enumerate}

% {\color{red} MFKeskin'in makalelerini (proceeding, direct/two step olan ve comparaitve eval olan) incele.}

\section{AoA-Sensing Vehicular VLC Receiver}

This section first presents the novel AoA-sensing VLC RX design, i.e., QRX, and then describes the AoA measurement procedure. The design promises high-rate communication and high-accuracy, high-resolution and high-rate AoA measurement, simultaneously, and is also low-cost since it considers COTS components only, enabling the practical realization of the restriction-free VLC-based vehicle localization solution.

\subsection{Receiver Design}

A conceptual diagram of the QRX and its optical configuration are shown in Fig. \ref{qrx_abstract} and a prototype of the QRX built by our group with low-cost COTS components is shown in Fig. \ref{qrx_proto}. The design of the QRX is inspired by the MIMO VLC RX in \cite{wang_hemisphericalAnalyz} but the QRX is specifically designed for high-resolution AoA measurement in VLP rather than simply achieving angular diversity. The QRX contains a hemispherical lens placed at a certain distance above a quadrant photodiode (QPD), converging the rays from the TX LED into a defocused spot. The spatial irradiance distribution on the QPD due to the spot, which depends on AoA by the ray optics relations that define $\lambda_{i}(\theta_{ij})$ in Eqn. (\ref{rx_model_2}) for each quadrant, determines the received signal power on each quadrant as depicted in Fig. \ref{qrx_abstract}. Let $f_{QRX}$ be the function that relates the AoA from TX $j$ to QRX $i$, i.e., $\theta_{ij}$, to the signal power ratio for signal $s_j$ from TX $j$ between horizontally separated quadrants, $\Phi_{ij}$, by

\vspace{-3mm}
\begin{equation}
\label{fqrx}
\Phi_{ij} = f_{QRX}(\theta) =  \frac{(\epsilon_{ij,B} + \epsilon_{ij,D})-(\epsilon_{ij,A} + \epsilon_{ij,C})}{\epsilon_{ij,A} + \epsilon_{ij,B} + \epsilon_{ij,C} + \epsilon_{ij,D}}~,
\end{equation}
\vspace{-2mm}

\noindent where $\epsilon_{ij,q}$ represents the power of the received signal in quadrant ${q, q \in \{A,B,C,D\}}$ of QRX $i$, and $\Phi_{ij}$ is bound to the [-1, 1] interval by definition. Choosing an $f_{QRX}$ function is the main task in QRX design since the inverse of $f_{QRX}$, which we call $g_{QRX}$, is used for measuring $\theta_{ij}$; $g_{QRX}$ is computed to sufficient precision by ray optics simulations offline and is stored in the form of a 1D look-up table on the ego vehicle.

% and can be calculated from quadrant received signal power readings online

$f_{QRX}$ is determined by the size of the spot and the range of its displacements from the QPD center due to non-zero $\theta_{ij}$: Both are determined by the optical configuration parameters, i.e., lens diameter, lens refractive index, QPD size and lens-QPD distance, denoted by $d_{L}$, $n$, $d_{H}$, $d_{X}$, respectively, as also shown in Fig. \ref{qrx_abstract}. $d_{L}$, $n$ and $d_{X}$ determine the size of the spot denoted by $d_S$, the full-width at half-maximum (FWHM) of its intensity distribution, which is trivially computed by \cite{edmund}:

\vspace{0mm}
\begin{equation}
	\label{dS}
	d_S = \frac{(d_{L})(d_{L}/n - d_{X})}{d_{L}/n} = d_{L}-(n)(d_{X})
\end{equation}
\vspace{-1mm}

\noindent since the distribution is approximately uniform. This approximation holds because 1) the QRX subtends a very small portion of the total non-uniform TX beam, thus, the beam arriving at the lens is truncated to an approximately uniform (i.e., \quotes{flat-top}) intensity distribution \cite{hurey}, and 2) the flat-top distribution experiences negligible diffraction artifacts after the lens since $d_X$ is significantly smaller than $d_L / n$ \cite{ozaktas}. Moreover, $d_{T}$, i.e., the displacement of the spot from the QPD center for a given $\theta_{ij}$, is determined by $d_{X}$ alone as follows: 

\begin{figure}[!t]
	\centering
	\includegraphics[width=0.95\linewidth]{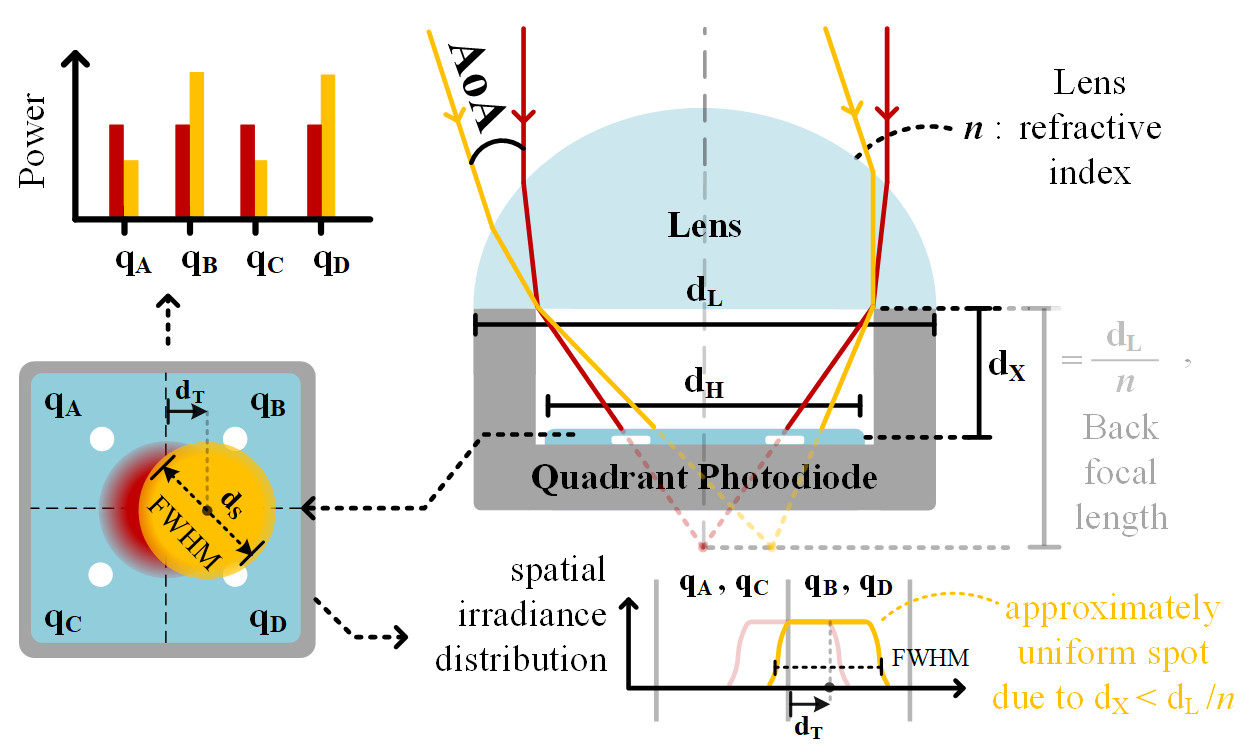}
	\vspace{2mm}
	\caption{Diagram of the AoA-sensing VLC RX, i.e., QRX. The red and yellow colors represent conditions for zero and non-zero AoA, respectively.}
	\label{qrx_abstract}
	\vspace{2mm}
	%\end{figure}
	%\begin{figure}[!t]
	%	\vspace{1mm}
	%	\centering
	\includegraphics[width=0.90\linewidth]{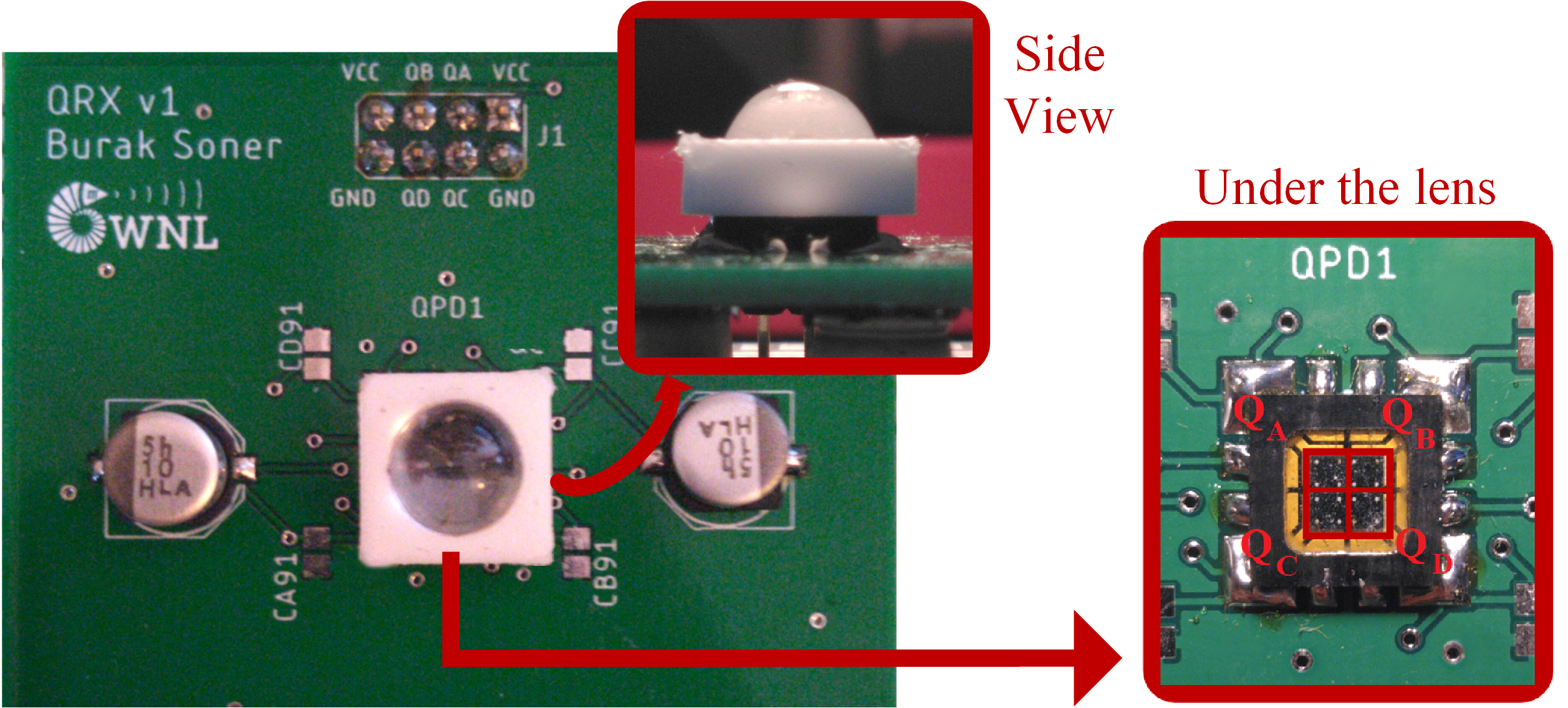}
	\vspace{2mm}
	\caption{Picture of the QRX prototype.}
	\label{qrx_proto}
	\vspace{-5mm}
\end{figure}

\vspace{-1mm}
\begin{equation}
	\label{dT}
	d_T = d_X \tan \left(\theta_{ij}\right)~.
\end{equation}
\vspace{-3mm}

\begin{figure}[t]
	\centering
	\includegraphics[width=.99\linewidth]{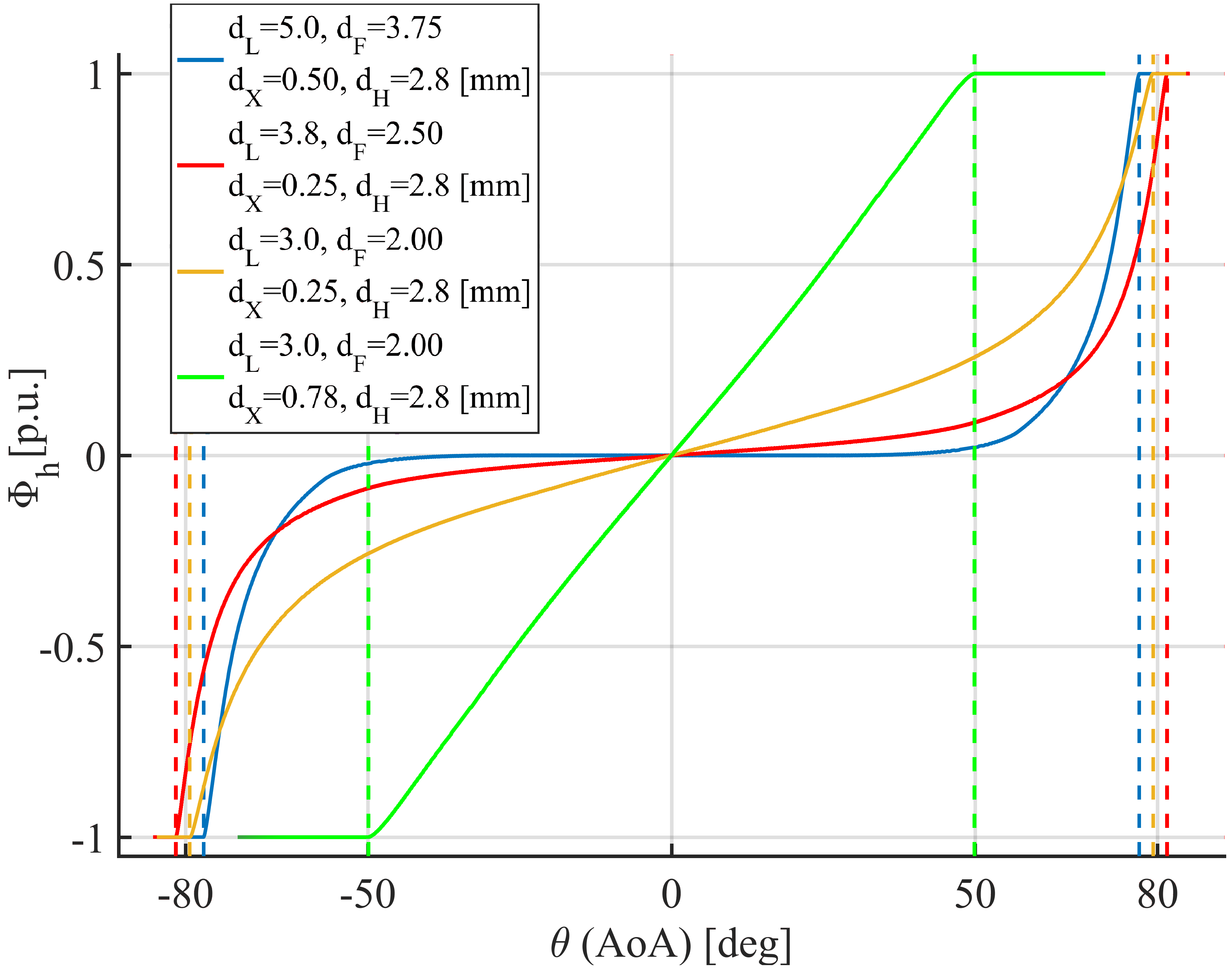}
	\vspace{-2mm}
	\caption{$f_{QRX}$ for different example configurations. Yellow curve provides the best compromise for the three design goals: bijection, high FoV and minimal non-linearity.}
	\label{qrx_lut}
	\vspace{-2mm}
\end{figure}

\noindent The relative magnitudes of $d_{S}$ and $d_{H}$, and the range of $d_{T}$ values that allow $\theta_{ij}$ measurement for a given configuration determine the conformance of $f_{QRX}$ to its design goals, which are: high FoV (ideally $\pm90$\textdegree), high linearity, and bijection, i.e., being one-to-one and onto. The FoV, denoted by $\theta_{FoV}$, is equal to the maximum measurable $\theta_{ij}$ value which is determined by a spot displacement of $d_T \! = \! d_S/2$ as per Eqns. (\ref{dS}) and (\ref{dT}), i.e., $\theta_{FoV} = \pm~\arctan(d_S/(2d_X))$, since all $\theta_{ij}$ outside the $\pm~\theta_{FoV}$ interval result in $|\Phi_{ij}| \! = \! 1$ due to two of the four quadrants receiving zero signal power, which renders $\theta_{ij}$ measurement impossible. Hence, larger $d_S$ or smaller $d_X$ provides larger FoV, however, this degrades linearity \cite{carbonneau1986optical}. Furthermore, ${d_{S} > d_{H}\sqrt{2}}$ violates bijection since in such configurations, all quadrants remain completely within the effective spot area and receive equal signal power for $|d_T| \approx 0$ \cite{manojlovic_qpdSns}, i.e., $\theta_{ij}$ values around zero become undetectable since they all result in ${\Phi_{ij} \! = \! 0}$, as in the blue curve in Fig. \ref{qrx_lut}. Therefore, recognizing these trade-offs, 1) a lens-QPD pair, i.e., \{$d_{L}, n, d_{H}$\}, and \linebreak 2) the lens-QPD distance, i.e., $d_{X}$, should be chosen to obtain the best compromise for the three design goals. 

\vspace{2mm}
\subsubsection{Choosing a lens-QPD pair}
The primary objective while choosing a lens-QPD pair is ensuring bijection. First, a low-cost and high-bandwidth COTS QPD is chosen to set $d_{H}$. Then, a lens is chosen to set $d_{L}$ and $n$ such that ${d_{S} < d_{H}\sqrt{2}}$ is ensured: Since $\{d_{L},d_{X},d_{S},n\}>0$ in Eqn. (\ref{dS}), $d_{L}$ upper-limits $d_{S}$, thus, setting $d_{L}<d_{H}\sqrt{2}$ guarantees bijection alone and makes $n$ a free parameter. Hence, a large $d_{L}$ that satisfies $d_{L}<d_{H}\sqrt{2}$ is chosen to also avoid constraining $d_S$, thus, the FoV, and $n$ is chosen solely with regards to low cost.

\subsubsection{Choosing the lens-QPD distance}
After setting \{$d_{L}, n, d_{H}$\}, $d_{X}$ is chosen to set $d_{S}$ as per Eqn. (\ref{dS}) for the best compromise between FoV and linearity. ${d_{H} < d_{S}}$ provides the highest FoV but the mapping is highly non-linear (red curve in Fig. \ref{qrx_lut}). $d_{S} \approx d_{H}$ still results in high FoV {($\approx$$\pm80$\textdegree)} and milder non-linearity (yellow curve in Fig. \ref{qrx_lut}). While ${d_{S} < d_{H}}$ linearizes the full dynamic range (green curve in Fig. \ref{qrx_lut}), this radically decreases the FoV \cite{carbonneau1986optical, crlbApertureRX}, thus, is not desirable. Therefore, ${d_{X}\approx (d_{L}-d_{H})/n}$ is chosen since {$d_{S} \approx d_{H}$} provides the best compromise, where $d_{L} < d_{H}\sqrt{2}$ to ensure bijection.

\subsection{Angle-of-Arrival Measurement}

\begin{figure*}[b]
	\centering
	\vspace{1mm}
	\includegraphics[width=.99\linewidth]{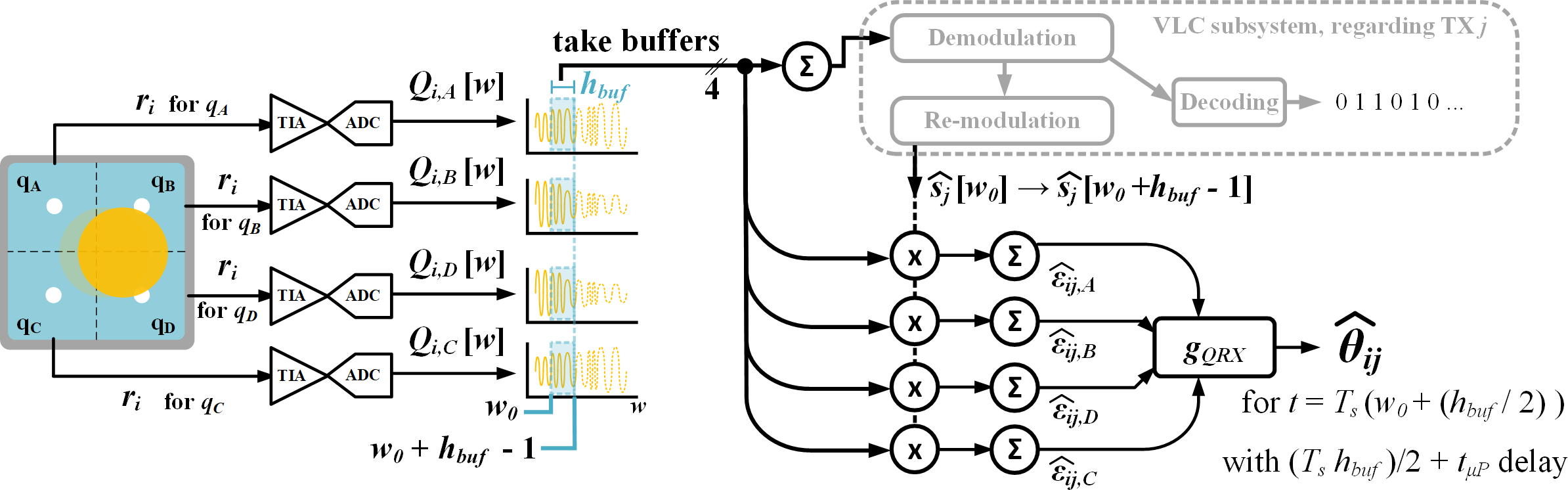}
	\vspace{1mm}
	\caption{Measurement procedure for AoA of TX $j$ to QRX $i$. Photocurrent received signals at each quadrant, $r_{i,q}, ~q \in \{A,B,C,D\}$, are converted to voltage signals via TIAs and sampled via ADCs to obtain the quadrant readings $Q_{i,q} [w]$, for sample time $w$. An $h_{buf}$ number of buffered samples are used for de/re-modulating the VLC TX signal to get $\widehat{s_j}[w]$, estimating $\epsilon_{ij,q}$, i.e., obtaining $\widehat{\epsilon_{ij,q}}$, and then for computing the AoA measurement, $\widehat{\theta_{ij}}$.}
	\label{qrx_algo}
	\vspace{-2mm}
\end{figure*}

For measuring AoA from TX $j$ to QRX $i$, first, the quadrant readings, i.e., the voltage signal produced by the TIA after amplifying $r_i$ for each quadrant $q$, are sampled at the Nyquist rate of the TX VLC waveform $s_j$, i.e., {1/$T_{s}$}, where $T_{s}$ is the sampling period. Then a number of $h_{buf}$ samples are buffered, which are used for both communication and AoA measurement purposes as follows: The VLC subsystem first demodulates the buffered noisy samples for obtaining the communication symbols. Then, it re-modulates the symbols to generate clean samples $\widehat{s_j}[w]$, which represent the contribution of $s_j$ in each buffer sample; $\widehat{s_j}[w]$ are used for estimating the signal power for $s_j$ on each quadrant of QRX $i$, i.e., $\epsilon_{ij,q}$, where $q \in \{A,B,C,D\}$, as $\widehat{\epsilon_{ij,q}}$ by:

\vspace{-3mm}
\begin{equation}
	\label{epsilon_i}
	\widehat{\epsilon_{ij,q}} = \frac{1}{h_{buf}}\left( \sum_{w=w_{0}}^{w_{0}+h_{buf}-1} \left(Q_{i,q}[w]\right)\left(\widehat{s_j}[w]\right) \right)~,
\end{equation}
\vspace{0mm}

\noindent where $w_{0}$ marks the sample time at the beginning of the buffer, and $Q_{i,q}[w]$ is the reading sample in quadrant $q$ at time ${w T_{s}}$. The estimations $\widehat{\epsilon_{ij,q}}$ are used for obtaining the AoA measurement, i.e., $\widehat{\theta_{ij}}$, as follows:

\vspace{0mm}
\begin{equation}
	\label{aoa_meas}
	\widehat{\theta_{ij}} = g_{QRX} \left(\frac{(\widehat{\epsilon_{ij,B}} + \widehat{\epsilon_{ij,D}})-(\widehat{\epsilon_{ij,A}} + \widehat{\epsilon_{ij,C}})}{\widehat{\epsilon_{ij,A}} + \widehat{\epsilon_{ij,B}} + \widehat{\epsilon_{ij,C}} + \widehat{\epsilon_{ij,D}}}\right)~.
\end{equation}
\vspace{0mm}

\noindent Eqns. (\ref{epsilon_i}) and (\ref{aoa_meas}) provide successful AoA measurement based on the following principle: Since $Q_{i,q}[w]$ are samples for $r_i$ which consist of the signal component and zero-mean AWGN as per Eqn. (\ref{rx_model_1}), the product $(Q_{i,q}[w])(\widehat{s_j}[w])$ results in a factor of the actual signal power scaled by the channel gain and contaminated by AWGN where the channel gain is not known. However, since Eqn. (\ref{aoa_meas}) considers a ratio of the estimated signal powers for each quadrant and the ratio values are tabulated for different AoA values, AoA measurement is possible even without knowing the exact channel gain as long as the ratio values are unique (i.e., $f_{QRX}$ is bijective), as described extensively in Section III-A. Furthermore, note that this measurement procedure does not dictate shape or bandwidth limitations for $s_j$, thus, does not impose any restrictive requirements. The only requirement, as also denoted in assumption (\textit{A2}), is that the VLC subsystem successfully demodulates the received signal and generates samples $\widehat{s_j}[w]$.

Overall, this procedure, depicted in Fig. \ref{qrx_algo}, produces AoA measurements for the buffer mid-points, i.e., for time ${t \! = \! T_s(w_{0} \! + \! (h_{buf})/2)}$ due to the non-weighted averaging operation in Eqn. (\ref{epsilon_i}), and the measurement rate is determined by $h_{buf}$, i.e., $f_u$ = 1/($T_{s} \cdot h_{buf}$), since $w_{0}$ is incremented by $h_{buf}$ for consecutive measurement cycles. However, since the time at which a measurement becomes available is ${t \! = \! T_s(w_{0} \! + \! h_{buf}) \! + \! t_{\mu P}}$ due to buffering and a finite processing time of $t_{\mu P}$, the measurement has a fixed delay of $(T_{s} \cdot h_{buf})/2 \! + \! t_{\mu P}$. Despite this delay, the AoA measurement can be done in real-time: The first term, $(T_{s} \cdot h_{buf})/2$, is negligible when rate is higher than 50 Hz as discussed in Section II-B. The second term, $t_{\mu P}$, which relates to the computational complexity of the procedure, is computed by:

\vspace{-4mm}
\begin{multline}
	\label{t_up}
	t_{\mu P} = \overbracket{~T_{FP}\left(k_\alpha h_{buf} \left( \log_2 (h_{buf})\right) + k_\beta \right)}^{t_{VLC}} \\ + \overbracket{~T_{FP}\left(2h_{buf} + h_{LU} \right)}^{t_{VLP}}~,
\end{multline}
\vspace{-2mm}

\noindent where $t_{VLC}$ is VLC demodulation and re-modulation time, ($k_\alpha$, $k_\beta$) are scalars to account for implementation-specific variations of the FFT-based modulation complexity, $t_{VLP}$ is the VLP processing time, $h_{LU}$ is the number of operations required for the $g_{QRX}$ table look-up, and $T_{FP}$ is the processor clock period. Considering modern processor speeds, known FFT-based modulation techniques \cite{comp_cplx}, and a $T_s$ of 1 $\mu$s and 50 Hz rate in the worst case (i.e., $h_{buf} = 1/(50\cdot10^{-6}) = 20.000$)~, $t_{\mu P}$ is only around a few $\mu$s, which is also negligible.

\begin{figure}[b]
	\centering
	\vspace{-2mm}
	\includegraphics[width=.99\linewidth]{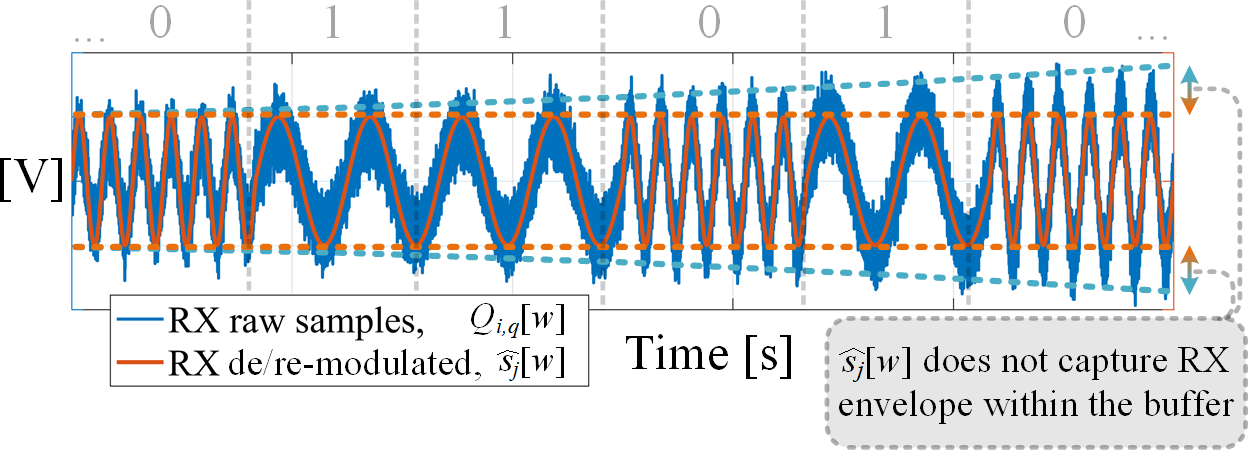}
	\vspace{-2mm}
	\caption{Exaggerated depiction of the \quotes{envelope effect} on $\widehat{s_j}[w]$ and $Q_{i,q}[w]$ for a single buffer with binary frequency shift keying modulated samples as an example; this effect, which causes inexact AoA measurement, is negligible for the targeted rates higher than 50 Hz.}
	\label{qrx_sjw}
\end{figure}

The accuracy of the AoA measurements is affected by two main factors: AWGN on the quadrant reading samples $Q_{i,q}[w]$, and the integrity of the samples $\widehat{s_j}[w]$ generated by the VLC subsystem. Since $\widehat{s_j}[w]$ does not have exact information on $H_{ij}$, it cannot track the RX signal envelope within the buffer, which means that considering Eqn. (\ref{epsilon_i}), $\widehat{\epsilon_{ij,q}}$ estimation, thus, AoA measurement, may not be exact for a non-constant signal envelope throughout an estimation cycle; this is depicted with exaggeration in Fig. \ref{qrx_sjw}. However, this effect, which is due to target vehicle movement within the buffer time interval, never becomes significant in practice since even the fastest vehicle transients, which are around 50 ms because of high inertia \cite{veh_dyn_Hzlevel}, do not fit within a single buffer time interval, which needs to take less than 20 ms considering the greater than 50 Hz rate requirement. Therefore, AoA measurements are contaminated predominantly by the AWGN on the VLC channel.

\section{Vehicle Localization with AoA-based VLP}

This section first presents the AoA-based VLP algorithm that is used for vehicle localization and then derives the associated Cramer-Rao lower bound (CRLB) on positioning accuracy which represents the sensitivity of the dual-AoA geometry used by the algorithm to errors in AoA measurements.

\subsection{Algorithm Description}
\begin{figure}[t]
	\centering
	\includegraphics[width=.75\linewidth]{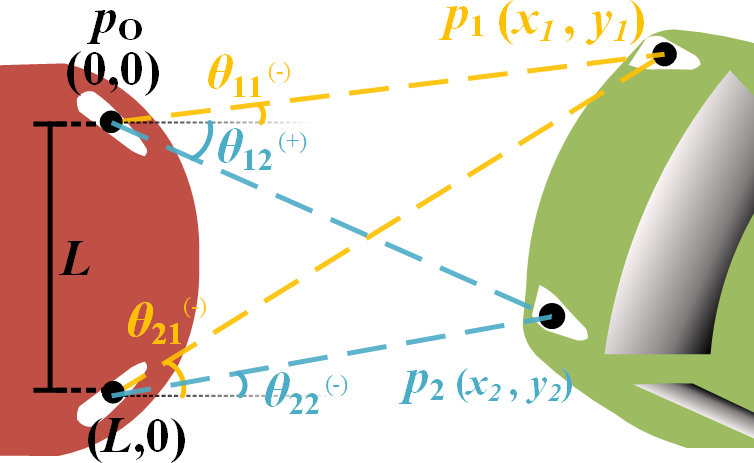}
	\vspace{1mm}
	\caption{The triangulation geometry used for dual-AoA-based VLP.}
	\label{aoa_vlp}
	\vspace{-2mm}
\end{figure}

The algorithm computes the relative position of two target TX units which define the vehicle location, via triangulation: TXs are located at the apexes of the triangles defined by two AoA measurements from two QRXs and the distance $L$ between them. Specifically, the position estimation for TX $j$, $\widehat{p_j}$, where $j \in \{1,2\}$, is computed by using $\widehat{\theta_{1j}}$, $\widehat{\theta_{2j}}$ and $L$, and the law of sines, as follows:

\vspace{-3mm}
\begin{equation}
\label{p_j}
\widehat{p_{j}} = 
\left[  
\begin{matrix}
\widehat{x_j}\\
\widehat{y_j}
\end{matrix}
\right] = 
\left[ 
\begin{matrix}
L\left(1 + \frac{\sin(\widehat{\theta_{2j}})~\times~\cos(\widehat{\theta_{1j}})}{\sin(\widehat{\theta_{1j}} - \widehat{\theta_{2j}})} \right)\\
L\left(\frac{\cos(\widehat{\theta_{2j}})~\times~\cos(\widehat{\theta_{1j}})}{\sin(\widehat{\theta_{1j}} - \widehat{\theta_{2j}})} \right)
\end{matrix}
\right]~,
\end{equation}
\vspace{-1mm}

\noindent and the localization rate is equal to the AoA measurement rate. The geometry for Eqn. (\ref{p_j}) is depicted in Fig. \ref{aoa_vlp}.

% \vspace{2mm}
% \subsubsection{Finding the Target Vehicle Pose}
% After finding the relative position of one target vehicle \ac{TX} unit, the calculation is repeated for the other \ac{TX} unit that faces the ego vehicle \ac{QRX}s. These two positions, i.e.,  $k_{tx1} = ({}^{e}x_{tx1}, {}^{e}y_{tx1})$ and $k_{tx2} = ({}^{e}x_{tx2}, {}^{e}y_{tx2})$, determine the relative pose of the target vehicle as per Eqns. (\ref{varphi_conv}) and (\ref{translt_conv}). 

\subsection{Cramer-Rao Lower Bound}

The CRLB is a bound on the mean-squared-error (MSE) for an unbiased estimate of a parameter given noisy observations that relate to that parameter through a given deterministic system model. Let $\mathbf{P}$ be a set of $N_{M}$ parameters in a system, let $\mathbf{M}$ be $N_{H}$ observations relating to that parameter, let $\mathbf{W}$ be the zero-mean AWGN terms with variance $\sigma_{\bf{W}}^2$ that contaminate those observations, and let $\mathbf{G}$ be the deterministic system model equations that relate the parameters and the observations. This observation model can be expressed as:

\vspace{-1mm}
\begin{equation}
	\label{crlb_sys}
	M_h = G_{h}(\mathbf{P}) + W_h~,~h=0,1,...N_H~~,
\end{equation}
\vspace{-3mm}

\noindent where $h$ is the index for elements $M_h$, $G_h$ and $W_h$ of vectors $\mathbf{M}$, $\mathbf{G}$ and $\mathbf{W}$, respectively. Since $W_h$ are AWGN, $M_h$ are independent Gaussian random variables with mean $G_{h}(\mathbf{P})$ and variance $\sigma_{W_h}^2$. Considering Eqn. (\ref{crlb_sys}), let $\widehat{\mathbf{P}}$ be the unbiased estimation of the parameter vector $\mathbf{P}$, and $\widehat{P_{m}}$ be the $m$\textsuperscript{th} element of that estimated parameter vector, $m \in \{1,2,...N_{M}\}$. The MSE in $\widehat{P_{m}}$ is lower bounded by 

\vspace{-1mm}
\begin{equation}
	\label{crlb_eqn}
	var(\widehat{P_m}) \geq {\left({\mathbf{F}^{-1}}\right)}_{m,m}~,
\end{equation}
\vspace{-3mm}

\noindent where $\bf{F}$ is the Fisher information matrix (FIM) and ${\left({\mathbf{F}^{-1}}\right)}_{m,m}$ denotes the $m$ by $m$ diagonal element of the inverse of the FIM. An estimator that satisfies Eqn. (\ref{crlb_eqn}) with equality is said to be efficient, i.e., no unbiased estimator that can provide smaller variance exists for the given problem. \linebreak The FIM is defined as

%and the right hand side corresponds to the diagonal elements of its inverse. While CRLB defines a full covariance matrix as the inverse of the FIM in its fundamental form [ref], we are interested only in the diagonal elements, which define the individual lower bound variances for each coordinate of the TX positions. While the general definition of the $N_m\times N_m$ FIM is 

\vspace{-3mm}
\begin{equation}
	\label{crlb_fim}
	\mathbf{F} = E[~\left(\nabla_{\mathbf{P}} \ln(p(\mathbf{M}|\mathbf{P})) \right)\left(\nabla_{\mathbf{P}} \ln (p(\mathbf{M}|\mathbf{P})) \right)^T~]~,
\end{equation}
\vspace{-3mm}

\noindent where $p$ is likelihood, $\nabla_{\mathbf{P}}$ denotes gradient with respect to $\mathbf{P}$, and $E$ denotes expectation. Since $M_h$ are independent Gaussian random variables, the expression for elements $(m,m')$ of the FIM simplifies to \cite[Ch. 3.9]{kay}:

\vspace{-2mm}
\begin{equation}
	\label{crlb_fim_simple}
	\small
	\mathbf{F}_{m,m'} = -\sum\limits_{h=1}^{N_h} \frac{1}{\sigma_{W_h}^2} \left( \frac{\delta G_{h}(\mathbf{P})}{\delta P_{m}} \cdot ¨\frac{\delta G_{h}(\mathbf{P})}{\delta P_{m'}} \right)~,
\end{equation}
\vspace{-1mm}

\noindent where $m, m' \in \{1,2,..., N_M\}$ and the FIM is $N_M \times N_M$. 

Based on this definition, the CRLB on positioning accuracy for the geometry used by the VLP algorithm can be derived with respect to the noisy AoA measurements: \linebreak $\mathbf{P} = \left[ x_1, y_1, x_2, y_2 \right]$ (i.e., $N_{M} \! = \! 4$), $\mathbf{M} = \left[ \widehat{\theta_{11}}, \widehat{\theta_{12}}, \widehat{\theta_{21}},  \widehat{\theta_{22}} \right]$ (i.e., $N_{H} \! = \! 4$), $\mathbf{G}$ is governed by Eqn. (\ref{rx_model_aoa}), and $\mathbf{W}$ is AWGN on $\widehat{\theta_{ij}}$ due to noise on the received signals used for the AoA measurements; the FIM is therefore 4x4. Note that this CRLB derivation 1) only captures accuracy against the AWGN on the VLC channel, thus, is only a measure of the \quotes{static} accuracy defined in Section II-B, 2) explicitly considers the sensitivity of the underlying geometric relations of the proposed dual-AoA VLP algorithm, i.e., it is a special case of the generic CRLB for multi-RX asynchronous VLP derived in \cite{mfkeskin_twostep}, and 3) assumes that the AWGN on the received signal propagates through to the AoA measurement: Since the AoA measurement expressions in Eqns. (\ref{epsilon_i}) and (\ref{aoa_meas}) are smooth, thus, piece-wise linear functions for reasonably small standard deviations of the AWGN-contaminated received signal around its expected value, $\widehat{\theta_{ij}}$ is also approximately a Gaussian random variable with mean $\theta_{ij}$ and variance $\sigma_{W_h}^2$. This phenomenon is thoroughly described in \cite[Ch. 3.6]{kay}, and it enables using Eqn. (\ref{crlb_fim_simple}) for the CRLB derivation. However, since the exact symbolic CRLB expression does not provide any extra intuition, only the derivative terms in Eqn. (\ref{crlb_fim_simple}) are presented here. The derivative expressions used for constructing the 4x4 FIM based on Eqns. (\ref{rx_model_aoa}) and (\ref{crlb_fim_simple}) are:

\begin{subequations}
	\vspace{0mm}
	\begin{equation}
		\label{crlb_aoa2_drv1}
		\frac{\delta \theta_{1j}}{\delta x_{1j}}= \frac{y_{1j}}{x_{1j}^2 + y_{1j}^2}~,~\frac{\delta \theta_{1j}}{\delta y_{1j}}= \frac{-x_{1j}}{x_{1j}^2 + y_{1j}^2}
	\end{equation}
	\vspace{0mm}
	\begin{equation}
		\label{crlb_aoa2_drv2}
		\small
		\frac{\delta \theta_{2j}}{\delta x_{1j}}= \frac{y_{1j}}{(x_{1j}-L)^2 + y_{1j}^2}~,~\frac{\delta \theta_{2j}}{\delta y_{1j}}= \frac{-(x_{1j}-L)}{(x_{1j}-L)^2 + y_{1j}^2}~,
	\end{equation}
	\vspace{-1mm}
\end{subequations}

\noindent where $j \in \{1,2\}$ and all other derivative terms are zero. The FIM is evaluated by using these derivative expressions during the simulations and the numerical value of the CRLB for a given condition (i.e., given $\mathbf{P}$ and $\mathbf{W}$), which represents the sensitivity of the dual-AoA geometry to errors in AoA measurement, is obtained as per Eqn. (\ref{crlb_eqn}) for comparison between theoretical and simulated performance.

\section{Simulations}

The simulations demonstrate the performance of the proposed VLC-based vehicle localization method under realistic road and VLC channel conditions in typical collision avoidance and platooning scenarios as well as comparing its simulated performance to the theoretical CRLB on localization accuracy with dual-AoA-based VLP. A custom MATLAB\textcopyright-based vehicular VLC simulator was built for this purpose, which is made available on GitHub \cite{github_link}. The simulator utilizes ego and target vehicle trajectories for different scenarios and generates the signals that emanate from the two target TXs and reach the two ego QRXs, for the whole trajectory, as per the system model equations presented in Section II-A. The proposed method first uses the simulated received signals to measure the AoA from the TXs to the QRXs by using the procedure proposed in Section III-B, and then provides relative localization based on these AoA measurements by using the algorithm described in Section IV-A. 

%The simulator is available on GitHub \cite{github_link}. 

Simulator setup parameters are given in Table \ref{table_sim}. The QRX design corresponds to the \quotes{best compromise} configuration, i.e., the yellow curve $f_{QRX}$ in Fig. \ref{qrx_lut}, but the photodiode dimensions in \cite{bechadergueConf_rtofPdoa} are used for fair comparison with existing results. In all simulations, the target vehicle leads the ego vehicle, thus, the target vehicle transmits through its tail light (similar to Fig. \ref{sys_mdl}). While the opposite configuration is equally valid, this configuration was chosen since it is the worst case scenario; the tail light has the lowest TX power among the vehicle lights. The simulated tail lights have 2 W optical power each and beam patterns are approximated by a Lambertian term of 20\textdegree ~half-power angle (order m=11) to be comparable to the setup in \cite{bechadergueConf_rtofPdoa}, which utilizes a total of two 2 W head lights and a 1 W tail light for VLP. The TX modulation scheme (binary frequency shift keying, BFSK) and bitrate complies with vehicle safety application requirements \cite{caileanSurvey_vlcAutoChlgs, VSCC_report}. A wide range of channel conditions, i.e., night-time versus day-time (indirect sunlight) light conditions, as well as clear versus foggy or rainy weather, are considered. Sunlight increases shot noise power as described in \cite{moreira1997optical}, and fog and rain attenuates the signal power as described in \cite{fograin}. We present four simulated driving scenarios to demonstrate that the proposed solution is eligible for collision avoidance and platooning applications under all of these channel conditions:

\begin{table}[!t]
	%\vspace{-3mm}
	\renewcommand{\arraystretch}{1.6}
	\caption{Simulator Setup Parameters}
	\label{table_sim}
	\centering
	\begin{tabular}{c|l|l}\hline
		\multirow{6}{*}{TX} & Signal & BFSK, $s_1, s_2$: 5/6, 12/13 kHz \\
		\cline{2-3}
		& Power &  $\gamma_j \cdot \max(\lvert s_j \lvert)$ = 2 W (tail light) \\
		\cline{2-3}
		& Pattern \cite{bechadergueThs_rtofPdoa} & Lambertian, m = $\floor*{\frac{-\ln2}{\ln(\cos(20\text{\textdegree}))}}$ = 11 \\
		\cline{2-3}
		& \multirow{3}{*}{Attenuation} & clear: - \\ 
		\cline{3-3}
		& & heavy rain ($\approx$10 mm/hr): 0.1 dB/m \cite{fograin}\\
		\cline{3-3}
		& & dense fog ($\approx$200 m):  0.3 dB/m \cite{fograin}\\
		\hline
		\hline
		\multirow{7}{*}{QRX TIA} & $\gamma_i$, $g_m$ & 0.5 A/W, 30 mS \\
		\cline{2-3}
		& $A_i$ (active) \textsuperscript{a} & 50 mm\textsuperscript{2} \\
		\cline{2-3}
		& $B$, $C_T$, $R_F$ & 10 MHz, 45 pF, 2.84 k$\Omega$, i.e., $G\approx10$ \\
		\cline{2-3}
		& Factors  & $\Gamma$=1.5 , $I_{B2}$=0.562 , $I_{B3}$=0.0868 \\
		\cline{2-3}
		& Temperature & $T$=298 K \\
		\cline{2-3}
		& \multirow{2}{*}{$I_{bg}$~\cite{moreira1997optical}} & night-time: 10 $\mu$A \\ 
		\cline{3-3}
		& & day, indirect sun: 750 $\mu$A\\
		\hline
		\hline
		\multirow{3}{*}{QRX Optics} & Lens & PMMA (n=1.5), $d_{L}$ = 7.1 mm\\
		\cline{2-3}
		& QPD  & $d_{H}$ = 6.3 mm , $d_{X}$ = 0.55 mm \\
		\cline{2-3}
		& FoV & $\pm50$\textdegree linear, $\pm80$\textdegree total (yellow, Fig. \ref{qrx_lut}) \\
		\cline{2-3}
		\hline
		\hline
		\multirow{2}{*}{Vehicle} & Dimensions & Length = 5 m. $L, D$ = 1.6 m \\
		\cline{2-3}
		& Steering & Ackermann \cite{ackermann} (small sideslip angles) \\
		\hline		
	\end{tabular}
	\justify \footnotesize \textsuperscript{a} Detection area is 50 mm\textsuperscript{2} in \cite{bechadergueConf_rtofPdoa} but converging/diverging optics usage is not specified. For fair comparison, QRX lens area, which is the detection area, is chosen as 50 mm\textsuperscript{2} here, thus, QPD area is 39.7 mm\textsuperscript{2} as per the design guidelines provided in Section III-A, and $C_T$ is scaled accordingly.
	\vspace{-3mm}
\end{table}

%%%%%%%%%%%%%%%%%%%%%%%%%%%%%%%%%%%%%%%%%%
% Simulation'daki SM1 -> text'teki SM3
% Simulation'daki SM2 -> text'teki SM1
% Simulation'daki SM3 -> text'teki SM2
% Simulation'daki SM4 -> text'teki SM4
% In order to justify SM3 assumptions for bechadergue fair comparison
%%%%%%%%%%%%%%%%%%%%%%%%%%%%%%%%%%%%%%%%%%

\vspace{1mm}
\begin{itemize}
	\item \textit{SM1} - dynamic, collision avoidance. A target vehicle leading an ego vehicle on a highway brakes dangerously during a lane change and risks collision. Results for different estimation rates are shown to demonstrate the effect of rate on localization accuracy for a highly dynamic target vehicle, as discussed in Section II-B. 
	\vspace{2mm}
	\item \textit{SM2} - dynamic, platooning. A target vehicle joins a platoon by moving in front of the ego vehicle from the left lane, drives on the same lane for a short while, and then exits the platoon towards the right lane. Results under all combinations of day-time and night-time light, and clear, foggy and rainy weather are shown for 100 Hz localization rate to demonstrate the performance of the method against adverse conditions for typical vehicle trajectories that occur during platooning.
	\vspace{2mm}
	\item \textit{SM3} - static, comparison with state-of-the-art (SoA) VLP. We simulate our method for the static vehicle locations described in \cite{bechadergueConf_rtofPdoa} under the same channel conditions. Results show that our method provides higher TX positioning accuracy for the high signal-to-noise ratio (SNR) regime under fair comparison, but \cite{bechadergueConf_rtofPdoa} is more resilient against low SNR. Additionally, the CRLB of the VLP algorithm is also evaluated for these locations to compare the theoretical and simulated performances.
	\vspace{2mm}
	\item \textit{SM4} - static, characterizing the operational range. This scenario considers the ego vehicle on the center of a 3-lane road and exhaustively simulates all feasible relative target locations under chosen favorable (i.e., night-time, clear) and challenging (i.e., day-time light, rain) channel conditions, characterizing the static accuracy of the method over its complete feasible operational range. 
\end{itemize}
\vspace{1mm}

\begin{figure*}[t]
	\centering
	\vspace{2mm}
	\includegraphics[width=.87\linewidth]{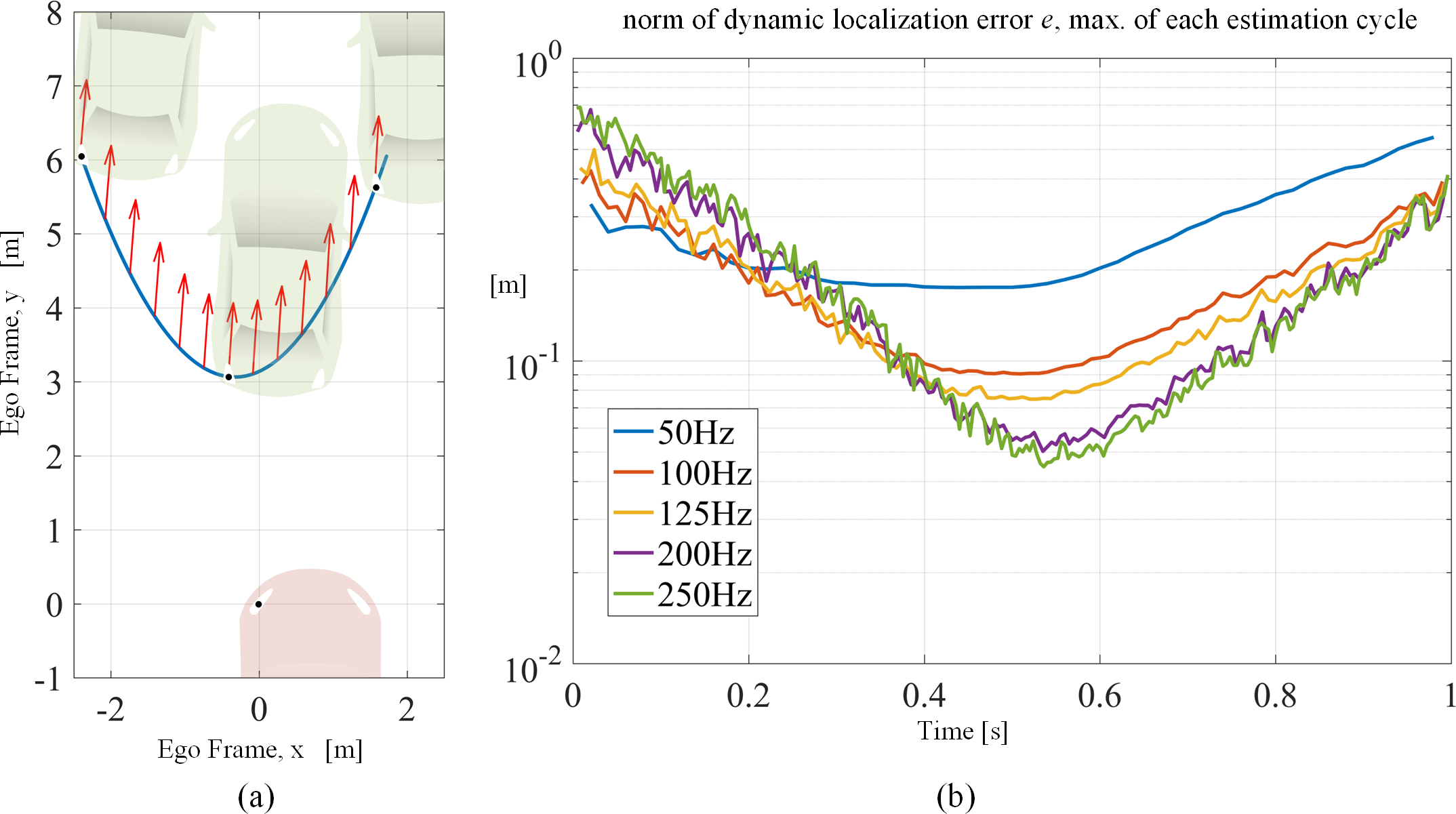}
	\vspace{2mm}
	\caption{\textit{SM1} - A typical collision avoidance scenario, dynamic vehicles. (a) Relative target vehicle trajectory. (b) Localization error, i.e., $\norm{e}$ as per Eqn. (\ref{accuracy}), over the trajectory that runs for a simulation time of 1 s, for different estimation rates under day-time (indirect sunlight) clear weather conditions.}
	\label{colAvd}
	\vspace{1mm}
\end{figure*}

\begin{figure*}[t]
	\centering
	\includegraphics[width=.93\linewidth]{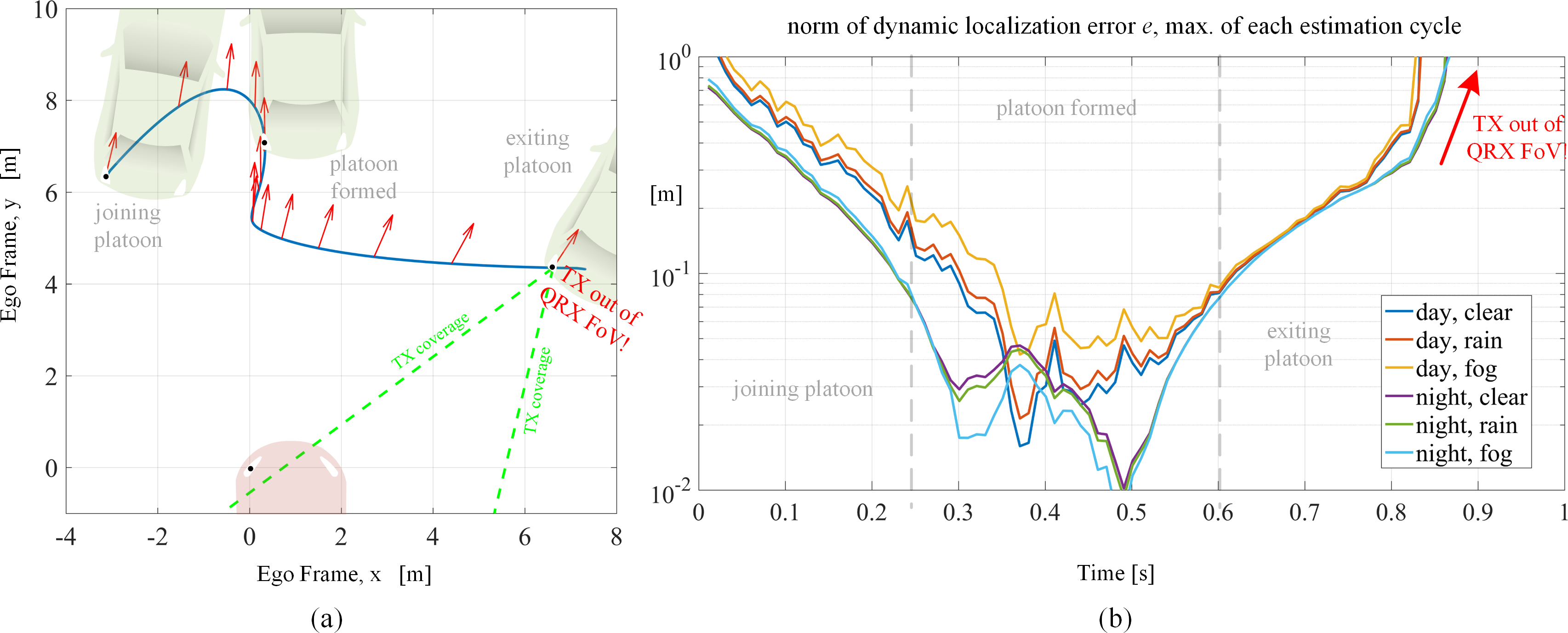}
	\vspace{1mm}
	\caption{\textit{SM2} - A typical platooning scenario, dynamic vehicles. (a) Relative target vehicle trajectory: Platoon formation, straight road platooning, and platoon dispersion over a simulation time of 1 s. (b) Localization error, i.e., $\norm{e}$ as per Eqn. (\ref{accuracy}), over the trajectory, for different weather and ambient light conditions.}
	\label{res_platoon}
	\vspace{-2mm}
\end{figure*}

\vspace{2mm}
\subsubsection{SM1 - Dynamic Scenario, Collision Avoidance}
Fig. \ref{colAvd} demonstrates the performance of the proposed method in a collision avoidance scenario under day-time, clear weather conditions for different estimation rates. Results show that cm-level accuracy is achieved for all rates greater than 100 Hz for the middle part of the trajectory, which has the maximum risk of collision. Furthermore, when the relative target vehicle movement is slow, decreasing the estimation rate increases accuracy in the case of low SNR; this can be seen in the results for the beginning of the trajectory where the target tail lights are facing away from the ego vehicle, decreasing received SNR, and lower rates provide better accuracy. However, lower rates decrease estimation accuracy when the target is highly dynamic due to the phenomenon described in Section II-B; this can be seen in the results for the middle of the trajectory. These results demonstrate that the proposed method provides cm-level accuracy for a typical collision avoidance scenario under high noise channel conditions, and estimation rate can be adjusted with respect to the SNR and the relative mobility of the vehicles for improving performance.

\vspace{2mm}
\subsubsection{SM2 - Dynamic Scenario, Platooning} 
Fig. \ref{res_platoon} demonstrates the performance of the proposed method for a platooning scenario (formation, straight road platooning, and dispersion) under clear, rainy and foggy weather conditions, and night and day ambient light for 100 Hz localization rate. Results show that while accuracy degrades severely due to sunlight and less severely due to fog and rain, cm-level accuracy is attained under all conditions for straight road platooning. Additionally, two practical irregularities of the proposed method are explicitly shown: 1) In the middle part of the trajectory, the SNR is very high and attenuation due fog and rain actually improves performance with respect to the clear weather case; this is because under night-time conditions, the signal power component in Eqn. (\ref{rx_noise_shot}) is the dominant noise source, which decreases significantly with attenuation. 2) Towards the end of the scenario, the TX units start pointing away from the QRX units and cause loss of estimation, demonstrating a practical limit of the proposed method due to the small angular coverage by tail lights. These results collectively demonstrate that the proposed method is capable of sustaining the accuracy and rate required for localization in a comprehensive platooning scenario under adverse weather and noise conditions, despite practical limitations.

\vspace{2mm}
\subsubsection{SM3 - Static Scenario, Comparison with SoA VLP} 

\begin{figure*}[t]
	\centering
	\includegraphics[width=.90\linewidth]{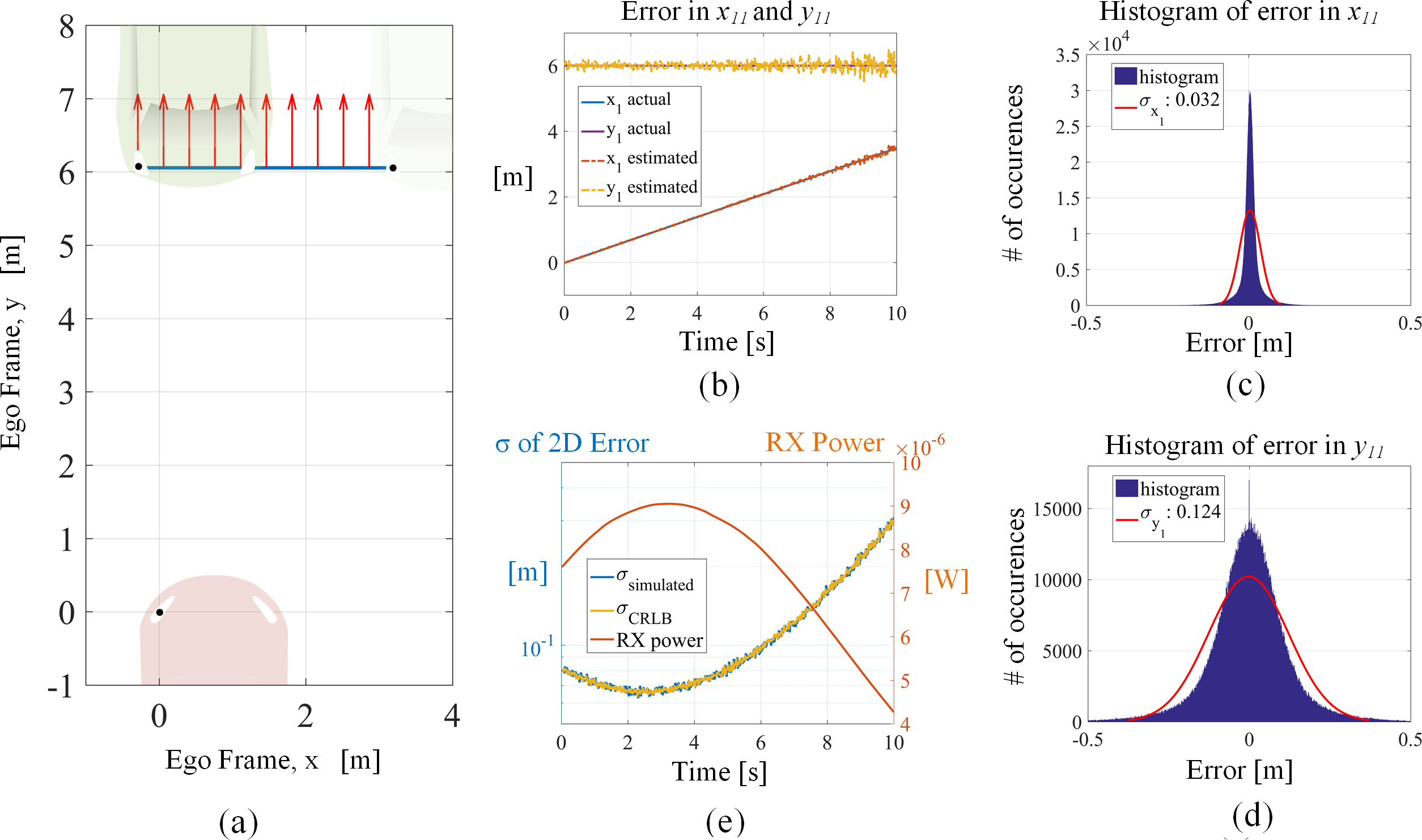}
	\vspace{1mm}
	\caption{\textit{SM3} - Comparison with SoA VLP. (a) Locations in \cite{bechadergueConf_rtofPdoa}: Vehicles are parallel and static at each simulation time step. (b) x and y estimation results for only $\widehat{p_1}$, under the same noise conditions in \cite{bechadergueConf_rtofPdoa}. (c) Localization error and the associated theoretical CRLB. Histogram of errors in (d) x, and (e) y.}
	\label{comparisonSoA}
	\vspace{-1mm}
\end{figure*}

This scenario demonstrates the performance of the proposed method for the static ego and target locations described in \cite{bechadergueConf_rtofPdoa} and depicted in Fig. \ref{comparisonSoA}a, under the same channel conditions. The following are addressed to ensure fair comparison: 

\vspace{1mm}
\begin{itemize}
	\item In \cite{bechadergueConf_rtofPdoa}, vehicles are assumed to stay parallel throughout this scenario, which involves the target vehicle slowly moving from 0 to 3.5 m of lateral distance: This does not define a realistic vehicle movement since a non-zero heading difference is required unless the target vehicle drifts sideways throughout the whole trajectory. Hence, we assume that the vehicles are static at each location, and separately simulate performance at each location. 
	\vspace{1mm}
	\item In \cite{bechadergueConf_rtofPdoa}, only results for a single TX on the target vehicle is provided. Therefore, we only consider the estimation of TX 1, i.e., $\widehat{p_1}$, for comparison, thus, the error vector in Eqn. (\ref{accuracy}) consists of $e_1$ only.
	\vspace{1mm}
	\item The method in \cite{bechadergueConf_rtofPdoa} is evaluated for 2 kHz rate, and our method is evaluated for 50 Hz. We do not evaluate our method at 2 kHz (or equivalently, \cite{bechadergueConf_rtofPdoa} at 50 Hz) since lower rate improves the static accuracy of our method but degrades that of \cite{bechadergueConf_rtofPdoa} due to heterodyning as explained in \cite{bechadergueThs_rtofPdoa}. We therefore evaluate both methods with their best reported configuration that is acceptable as per collision avoidance and platooning requirements.
\end{itemize}
\vspace{1mm}

Fig. \ref{comparisonSoA}b demonstrates the x and y estimation performances of the proposed method, and Figs. \ref{comparisonSoA}c and \ref{comparisonSoA}d show the histogram of associated errors over the trajectory, sampled over 1000 iterations for higher statistical significance. While \cite{bechadergueConf_rtofPdoa} provides higher accuracy in the y axis on average over the whole trajectory, i.e., 6.2 cm error in \cite{bechadergueConf_rtofPdoa} versus 12.4 cm error in ours, our method provides superior accuracy in the x axis, i.e., 11.3 cm error in \cite{bechadergueConf_rtofPdoa} versus 3.2 cm error in ours. The difference in x and y estimation performance for our method is due to the difference in the sensitivities of the sin/cos non-linearities in Eqn. (\ref{p_j}). In terms of overall 2D accuracy, our method provides better performance for the beginning of the trajectory, where the SNR is higher as shown in Fig. \ref{comparisonSoA}e, i.e., approximately 12.9 cm error in \cite{bechadergueConf_rtofPdoa} versus less than 10 cm error in ours. However, \cite{bechadergueConf_rtofPdoa} is more resilient against noise and sustains this performance for also the lower SNR regime at the end of the trajectory, contrary to our method. Additionally, the CRLB of our method is evaluated since static locations are considered. Simulated accuracy meets the CRLB as shown in Fig. \ref{comparisonSoA}e, demonstrating that the proposed VLP algorithm is an efficient estimator, i.e., the minimum variance unbiased estimator for the dual-AoA vehicular VLP problem. These results show that our proposed method advances the state-of-the-art for vehicular VLP with cm-level localization accuracy and 50 Hz rate under realistic road and channel conditions except for very low SNR. Furthermore, our proposed method does not impose any high-bandwidth circuit requirements like \cite{bechadergueConf_rtofPdoa} and is therefore feasible for general use.

\vspace{3mm}
\subsubsection{SM4 - Characterizing the Operational Range}

\begin{figure}[t]
	\centering
	\includegraphics[width=.93\linewidth]{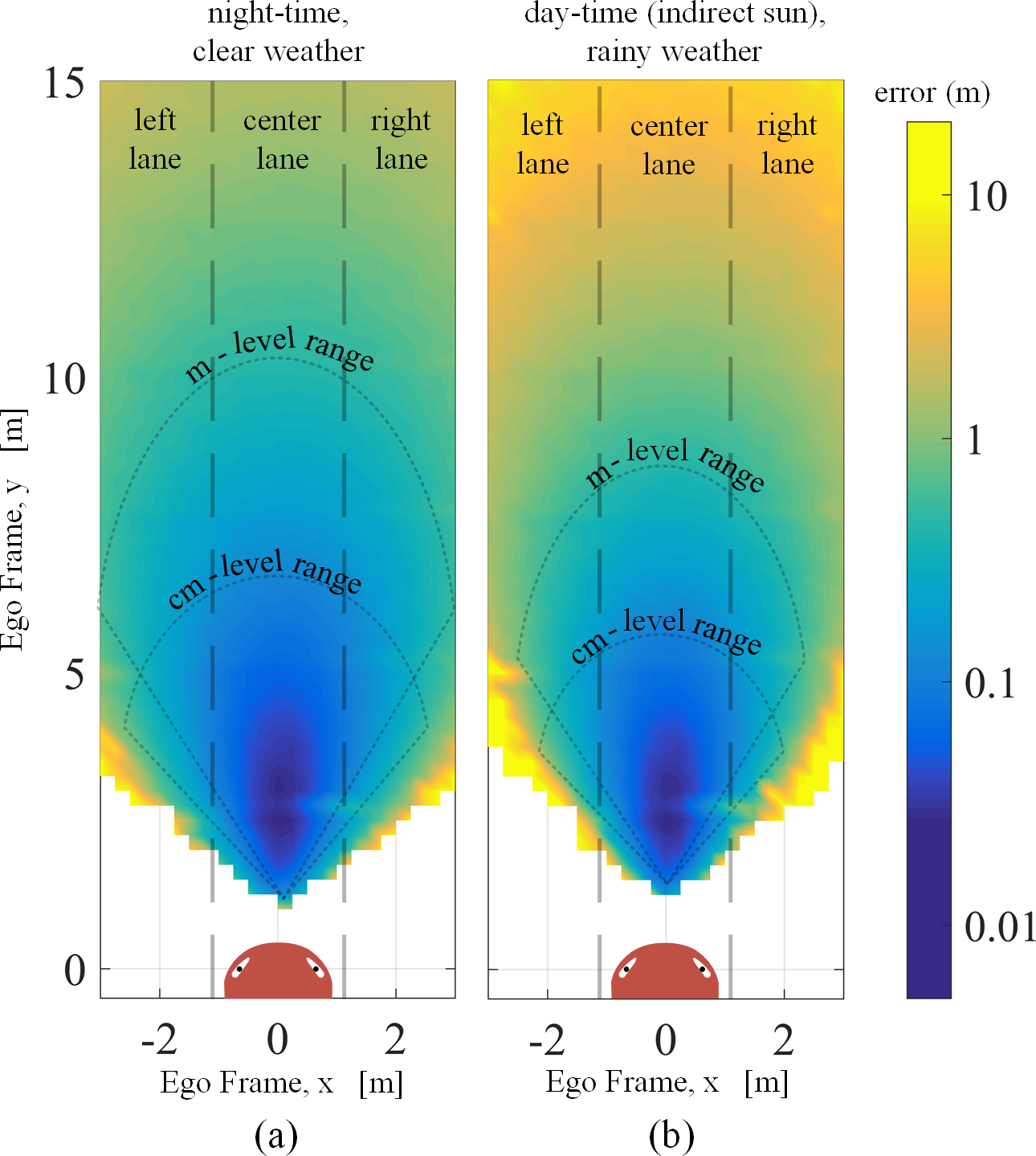}
	\vspace{1mm}
	\caption{\textit{SM4} – Characterizing the operational range for a 3-lane road scenario (static locations). Results for each location show average error over all feasible target orientations for 50 Hz localization rate, under (a) night-time, clear weather conditions, and (b) day-time, rainy weather conditions.}
	\label{exh_res}
	\vspace{-3mm}
\end{figure}

The procedure for this scenario is as follows: Estimation accuracy for sampled relative target vehicle locations over three lanes \linebreak ($\pm$3 m horizontal distance from ego vehicle bumper center) and 15 m longitudinal distance are evaluated to characterize the feasible operational range of the proposed method for collision avoidance and platooning scenarios, where the recorded performance for each location is an average of the results for all feasible target orientations at that location. The estimation rate is 50 Hz. Figs. \ref{exh_res}a and \ref{exh_res}b demonstrate localization accuracy under favorable and harsh conditions, i.e., night-time clear weather, and day-time heavy rain (10 mm/hr), respectively. The colorless zones in the graphs denote a loss of estimation for those locations due to QRX units being out of TX coverage, which occurs due to the target vehicle being either too close to the ego front bumper (less than 1 m), or at an angle too oblique to be considered as either a platoon element \cite{sartre} or a tangible collision threat \cite{destatis}. Therefore, these locations are not strictly relevant for collision avoidance and platooning scenarios. Fig. \ref{exh_res}a shows that for favorable channel conditions the promised cm-level accuracy is attained within approximately a 7 m radius. Outside the 7 m radius, up until the 10 m mark, accuracy is still better than 1 m. On the other hand, under harsh channel conditions, cm-level accuracy is limited to a radius of approximately 5 m, as shown in Fig. \ref{exh_res}b. The accuracy is still better than 1 m up to the 8 m mark. These results demonstrate that the proposed method provides cm-level accuracy with at least 50 Hz rate even under harsh channel conditions for configurations relevant for collision avoidance and platooning within approximately 10 m range.

\begin{table}[t]
	%\vspace{-3mm}
	\renewcommand{\arraystretch}{1.5}
	\caption{Comparison of vehicle localization methods}
	\label{table_comparison}
	\centering
	\begin{tabular}{c|c|c|c}
		Method & \# of ops / cycle & Error & Rate \\
		\hline
		\hline
		This paper & $\approx$ 10\textsuperscript{5} & $\leq$ 10 cm & $\geq$ 50 Hz\\
		\hline
		RToF-VLP \cite{bechadergueConf_rtofPdoa} & $\approx$ 10\textsuperscript{5} & $\leq$ 10 cm  & $\geq$ 50 Hz\\
		\hline
		RADAR \cite{radar} & $\approx$ 10\textsuperscript{7} & $\geq$ 10 cm  & $\approx$ 50 Hz\\
		\hline
		Camera \cite{cam_survey} & $\approx$ 10\textsuperscript{9} & $\approx$ 10 cm  & $\leq$ 50 Hz\\
		\hline
	\end{tabular}
	\vspace{-3mm}
\end{table}

Considering these results, a rough comparison of vehicle localization methods that considers the same distance range and FoV, in terms of computational complexity, accuracy and rate, is provided in Table \ref{table_comparison}. While RADAR-based methods can provide up to 10 cm accuracy, they have significantly high complexity, thus, typically low rate even with SoA computational optimizations, due to the large number of receiving elements used (typically more than 1000 antennas in phased-array form), as thoroughly described in \cite{radar}. Camera-based methods provide cm-level accuracy but suffer even more from the high number of receiving elements (i.e., typically more than a million pixels), thus, are fundamentally limited to low rates considering a feasible cost margin for automotive use \cite{cam_survey}. VLC-based methods are multiple orders of magnitude less complex than sensor-based methods since very few receiving elements, situated at known locations on the vehicles, are used, enabling real-time service; the complexity analysis of our proposed method provided in Eqn. (\ref{t_up}) showcases the simplicity of VLC-based methods. Furthermore, our proposed VLC-based method attains similar accuracy as its best alternative, the RToF-based VLP in \cite{bechadergueConf_rtofPdoa}, but without imposing requirements that restrict practical use as in \cite{bechadergueConf_rtofPdoa}.

\section{Conclusion}
This paper proposes a novel VLC-based vehicle localization method based on the design of a novel low-cost/size VLC receiver (\quotes{QRX}) that can simultaneously provide high-rate communication, and high-accuracy, high-resolution and high-rate AoA measurement. The QRX can be realized with COTS components only, enabling the first practical realization of an AoA-based vehicular VLP method with cm-level accuracy at greater than 50 Hz rate without imposing any restrictive requirements such as limited vehicle orientations and use of high-bandwidth circuits, localized road-side lights, and VLC waveform constraints. The VLP algorithm uses two QRXs to locate two target vehicle VLC TX units relative to the ego vehicle via VLP, which is sufficient for relative vehicle localization. The computational overhead of the solution is very low compared to the conventional sensor solutions. Performance of the proposed solution is evaluated with theoretical CRLB analysis and exhaustive simulations on a custom vehicular VLC simulator. Simulations demonstrate that the proposed solution performs localization with cm-level accuracy at greater than 50 Hz rate even under harsh road and VLC channel conditions. The solution is expected to complement the existing autonomous vehicle sensor system for higher safety by providing vehicle localization for collision avoidance and platooning applications. 

As future work, we plan to derive the CRLB for existing state-of-the-art VLP algorithms, implement the corresponding VLC-based localization methods in simulation environment, and provide fair comparisons of those methods and the method proposed in this paper under the same realistic driving scenarios and channel conditions. Furthermore, we plan to test the proposed method with hardware built by our group under real driving scenarios to experimentally verify the results reported in this paper.

% \section*{Acknowledgement}
% The authors acknowledge the support of Ford Otosan and the Scientific and Technological Research Council of Turkey EU CHIST-ERA grant \# 119E350.

\ifCLASSOPTIONcaptionsoff
  \newpage
\fi

\bibliographystyle{ieeetr}
\bibliography{trns1}
\nocite{}

\begin{comment}
\vspace{-15mm}
\begin{IEEEbiography}[{\includegraphics[width=1in,height=1.25in,clip,keepaspectratio]{"burak.jpg"}}]{Burak Soner}
Biography text here.
(Member, IEEE) received the B.Sc. degree in Mechatronics from Sabancı University, Istanbul, Turkey, in 2014. He is currently pursuing the Ph.D. degree in Electrical and Electronics engineering with Koç University (KU), Istanbul, Turkey. Until 2016, he worked on power electronics at Sabancı University Microsystems Lab as a Researcher and on automotive embedded control/safety systems at AVL List GmbH as a Systems Engineer. He worked on real-time computation and compression for holographic displays at CY Vision, San Jose, CA, USA, and at the Optical Microsystems Laboratory, KU, until 2019. He joined the Wireless Networks Laboratory, KU, in January 2019, where he is currently working on vehicular communications and navigation. His specific research interests are optical wireless communication and communication-based positioning/navigation systems for ground and aerial vehicles.
\vspace{-5mm}
\end{IEEEbiography}

%\newpage

\vspace{-10mm}

\begin{IEEEbiography}{Sinem Coleri}
Biography text here.
\end{IEEEbiography}
\end{comment}

\end{document}